\documentclass[12pt,fleqn,a4paper]{article} 
\usepackage{epsfig,graphics,graphicx}
\usepackage{clshan-math,clshan-dm-dd}
\def\lsim{\:\raisebox{-0.5ex}{$\stackrel{\textstyle<}{\sim}$}\:}

\begin{document}
\thispagestyle{empty}
\begin{flushright}
 March 2007
\end{flushright}
\begin{center}
{\Large\bf Reconstructing the Velocity Distribution of WIMPs \\ \vspace{0.2cm}
           from Direct Dark Matter Detection Data}           \\
\vspace*{7mm}
 {\sc Manuel Drees} and {\sc Chung-Lin Shan}                 \\
\vspace*{3mm}
 {\it Physikalisches Inst. der Univ. Bonn, Nussallee 12, 53115 Bonn, Germany}
\end{center}
\vspace*{10mm}
\begin{abstract}
  Weakly interacting massive particles (WIMPs) are one of the leading
  candidates for Dark Matter. Currently, the most promising method to detect
  many different WIMP candidates is the direct detection of the recoil energy
  deposited in a low--background laboratory detector due to elastic
  WIMP--nucleus scattering. So far the usual procedure has been to predict the
  event rate of direct detection of WIMPs based on some model(s) of the
  galactic halo. The aim of our work is to invert this process. That is, we
  study what future direct detection experiment can teach us about the WIMP
  halo. As the first step we consider a time--averaged recoil spectrum,
  assuming that no directional information exists. We develop a method to
  construct the (time--averaged) one--dimensional velocity distribution
  function from this spectrum. Moments of this function, such as the mean
  velocity and velocity dispersion of WIMPs, can also be obtained directly
  from the recoil spectrum. The only input needed in addition to a measured
  recoil spectrum is the mass of the WIMP; no information about the scattering
  cross section or WIMP density is required.
\end{abstract}
\clearpage
\section{Introduction}

The first indication of the existence of Dark Matter has already been found in
the 1930s \cite{evida}. By now astrophysicists have strong evidence
\cite{evida}-\cite{WMAP} to believe that a large fraction (more than 80\%) of
the matter in the Universe is dark (i.e., interacts at most very weakly with
electromagnetic radiation). The dominant component of this cosmological dark
matter must be due to some yet to be discovered, non--baryonic particles.
Weakly interacting massive particles (WIMPs) $\chi$ are one of the leading
candidates for Dark Matter. WIMPs are stable particles which arise in several
extensions of the standard model of electroweak interactions. Typically they
are presumed to have masses between 10 GeV and a few TeV and interact with
ordinary matter only weakly (for reviews, see \cite{susydm}).

Currently, the most promising method to detect many different WIMP candidates
is the direct detection of the recoil energy deposited in a low--background
laboratory detector by elastic scattering of ambient WIMPs on the nuclei in a
detector \cite{detaa}-\cite{detac}. The event rate of direct WIMP detection
depends strongly on the velocity distribution of the incident particles.
Usually and for simplicity, the local velocity distribution of WIMPs is
assumed to be a shifted Maxwell distribution, as would arise if the Milky Way
halo is isothermal \cite{detaa}-\cite{modela}. However, our halo is certainly
not a precisely isothermal sphere. Possibilities that have been considered in
the literature include axisymmetric halo models \cite{modelb}, the so--called
secondary infall model of halo formation \cite{modelc}, and a possible bulk
rotations of the halo of our galaxy \cite{modeld,annualab}.

If the halo of our galaxy consists of WIMPs, about $10^5$ WIMPs should pass
through every square centimeter of the Earth's (and our!) surface per second
(for $m_{\chi} \simeq 100~{\rm GeV}$). However, the cross section of WIMPs on
ordinary materials is very low and makes these interactions quite rare
\cite{susydm}.  For example, in typical SUSY models with neutralino WIMP, the
event rate is about $10^{-4} \sim 1$ event/kg/day and the energy deposited in
the detector by a single interaction is about $1 \sim 100$ keV. Typical event
rates due to cosmic rays and ambient radioactivity are much larger. The annual
modulation of the event rate due to the orbital motion of the Earth around the
Sun has been suggested as a way to discriminate signal from background
\cite{detab,detac,annualaa,annualab}. Actually, the DAMA collaboration has
claimed that they have observed this annual modulation of the event rate
\cite{DAMA}. Note, however, that the annual modulation of the signal is
expected to amount only to a few percent; this method can therefore only be
used once more than one hundred signal events have been accumulated. In the
meantime, a more promising approach is to reduce the background by vetoing
events that do not look like nuclear recoil. This method is, e.g., being used
by the CDMS \cite{CDMS}, CRESST \cite{CRESST} and EDELWEISS \cite{EDELWEISS}
collaborations. The presently best null result, from CDMS \cite{CDMSbound},
contradicts the DAMA claim for standard halo models.
\footnote{The two
 experiments might be compatible in some exotic scenarios. One possibility is
 to postulate rather light, $m_\chi < 10$ GeV, and fast WIMPs with large
 scattering cross section \cite{gelmini}. Another possible way out is to
 postulate that the detected events are actually inelastic, leading to the
 production of a second particle that is almost, but not exactly, degenerate
 with the WIMP \cite{inel}.}

So far most theoretical analyses of direct WIMP detection have predicted the
detection rate for a given (class of) WIMP(s), based on a specific model of
the galactic halo. The goal of this paper is to invert this process. That is,
we wish to study, as model--independently as possible, what future direct
detection experiments can teach us about the WIMP halo. In other words, we
want to start the (theoretical) exploration of ``WIMP astronomy''. In this
first study we use a time--averaged recoil spectrum, and assume that no
directional information exists. We can thus only hope to construct the
(time-averaged) one--dimensional velocity distribution $f_1(v)$, where $v$ is
the absolute value of the WIMP velocity in the Earth rest frame.
 Note that our ansatz is quite different from that of the recent paper \cite{Green07},
 which assumes a WIMP velocity distribution
 and then analyses with which precision the WIMP mass can be determined
 from the direct detection experiment.

The remainder of this article is organized as follows. In Sec.~2 we show how
to find the velocity distribution of WIMPs from the functional form of the
recoil spectrum; our assumption here is that this functional form has been
determined by fitting the data of some (future) experiment(s). We then derive
formulae for moments of the velocity distribution function, such as the mean
velocity and the velocity dispersion of WIMPs, which can be compared with
model predictions. We also discuss some simple halo models. In Sec.~3 we will
develop a method that allows to reconstruct the WIMP velocity distribution
function directly from recorded signal events. This allows statistically
meaningful tests of predicted distribution functions. We will also show how to
calculate the moments of the velocity distribution directly from these data.
In Sec.~4 we conclude our work and discuss some further projects. Some
technical details of our calculations are given in the Appendices.

\section{One--dimensional velocity distribution function}

In this section we first show how to reconstruct (moments of) the WIMP
velocity distribution, and then discuss some simple model distributions.

\subsection{Reconstructing the velocity distribution function}

The differential rate for elastic WIMP--nucleus scattering is given by
\cite{susydm}:
\beq \label{eqn201}
\dRdQ = \calA \FQ \intvmin \bfrac{f_1(v)}{v} dv\, .
\eeq
Here $R$ is the direct detection event rate, i.e., the number of events per
unit time and unit mass of detector material, $Q$ is the energy deposited in
the detector, $F(Q)$ is the elastic nuclear form factor, and $f_1(v)$ is the
one--dimensional velocity distribution function of the WIMPs impinging on the
detector. The constant coefficient $\calA$ is defined as
\beq \label{eqn202}
\calA \equiv \frac{\rho_0 \sigma_0}{2 m_{\chi} m_{\rm r}^2}\, ,
\eeq
where $\rho_0$ is the WIMP density near the Earth and $\sigma_0$ is the total
cross section ignoring the form fact suppression. The reduced mass $m_{\rm r}$ is
defined by 
\beq \label{eqn203}
m_{\rm r}  \equiv \frac{m_{\chi} m_{\rm N}}{m_{\chi}+m_{\rm N}}\, ,
\eeq
where $m_{\chi}$ is the WIMP mass and $m_{\rm N}$ that of the target nucleus.
Finally, $v_{\rm min}$ is the minimal incoming velocity of incident WIMPs that
can deposit the energy $Q$ in the detector:
\beq \label{eqn204}
v_{\rm min} = \alpha \sqrt{Q}\, ,
\eeq
where we define 
\beq \label{eqn205}
\alpha \equiv \sfrac{m_{\rm N}}{2 m_{\rm r}^2} .
\eeq
In Eqs.(\ref{eqn201})--(\ref{eqn205}) we have assumed that the detector
essentially only consists of nuclei of a single isotope. If the detector
contains several different nuclei (e.g. NaI as in the DAMA detector), the
right--hand side (rhs) of Eq.(\ref{eqn201}) has to be replaced by a sum of
terms, each term describing the contribution of one isotope. For simplicity,
in the remainder of this article we will focus on mono--isotopic detectors.

In this section we wish to invert Eq.(\ref{eqn201}), i.e., we want to find an
expression for the one--dimensional velocity distribution function $f_1(v)$
for given (as yet only hypothetical) measured recoil spectrum $dR/dQ$. To
that end, we first define
\beq
\Dd{F_1(v)}{v} = \frac{f_1(v)}{v} \, ,
\eeq
i.e., $F_1(v)$ is the primitive of $f_1(v)/v$.  Eq.(\ref{eqn201}) can then be
rewritten as
\beq \label{eqn211}
\frac{1}{\calA \FQ} \adRdQ = \intvmin \bfrac{f_1(v)}{v} dv
 = F_1(v \to \infty) - F_1(v_{\rm min}).
\eeq
Since WIMPs in today's Universe are quite slow, $f_1(v)$ must vanish as $v$
approaches infinity, 
\beq
f_1(v \to \infty) \to 0 \, .
\eeq
Hence
\beq
\Dd{F_1(v)}{v} \Bigg|_{v \to \infty} = \frac{f_1(v)}{v} \Bigg|_{v \to \infty}
 \to 0.
\eeq
This means that $F_1(v\to\infty)$ approaches a finite value. Differentiating
both sides of Eq.(\ref{eqn211}) with respect to $v_{\rm min}$ and using
Eq.(\ref{eqn204}), we find
\beqnN
\Dd{F_1(v_{\rm min})}{v_{\rm min}} \&= \&- \frac{1}{\calA} \left\{ \dd{v_{\rm
      min}} \bdRdQoFQ_{Q=v_{\rm min}^2/\alpha^2}\right\} 
       \nonumber\\
       \nonumber\\
 \&=\& \frac{1}{v_{\rm min}} \afrac{1}{\calA}
       \left\{-2 Q \cdot \ddRdQoFQdQ\right\}_{Q=v_{\rm min}^2/\alpha^2}.
\eeqnN
Since this expression holds for arbitrary $v_{\rm min}$, we can write down the
following result directly: 
\beq \label{en1}
\frac{f_1(v)}{v} = \Dd{F_1(v)}{v} = \frac{1}{v} \afrac{1}{\calA} \left\{-2 Q
  \cdot \ddRdQoFQdQ\right\}\Qva .
\eeq

The rhs of Eq.(\ref{en1}) depends on the as yet unknown constant
$\calA$. Recall, however, that $f_1$ is the {\em normalized} velocity
distribution, i.e., it is defined to satisfy
\beq \label{eqn214}
\intz f_1(v) \~ dv = 1 .
\eeq
Therefore, the normalized one--dimensional velocity distribution function can
be written as 
\beq \label{eqn212}
f_1(v) = \calN \cbrac{-2 Q \cdot \ddRdQoFQdQ}\Qva \, ,
\eeq
with normalization constant $\calN$ (see Appendix \ref{appN})
\beq \label{eqn213}
\calN = \frac{2}{\alpha} \cbrac{\intz \frac{1}{\sqrt{Q}} \bdRdQoFQ
  dQ}^{-1} \, .
\eeq
Note that the integral on the rhs of Eq.(\ref{eqn213}) starts at $Q=0$.

In the next step we wish to compute {\em moments} of the distribution function
$f_1$:
\beq \label{en2}
\langle v^n \rangle = \int_{v_{\rm min}(Q_{\rm thre})}^\infty v^n f_1(v) \~ dv
\, .
\eeq
In Eq.(\ref{en2}) we have introduced a threshold energy $Q_{\rm thre}$. This is
needed experimentally, since at very low recoil energies, the signal is
swamped by electronic noise. Moreover, we will later meet expressions that
(formally) diverge as $Q \rightarrow 0$. $v_{\rm min}(Q_{\rm thre})$ is
calculated as in Eq.(\ref{eqn204}). Inserting Eq.(\ref{eqn212}) into
Eq.(\ref{en2}) and integrating by parts, we find (see Appendix \ref{appN})
\beq \label{eqn215}
\langle v^n \rangle = \calN \afrac {\alpha^{n+1}} {2} \left[ \frac
    {2 \~ Q_{\rm thre}^{(n+1)/2}} {F^2(Q_{\rm thre})} \afrac{dR}{dQ}_{Q=Q_{\rm thre}}
 + (n+1) I_n(Q_{\rm thre}) \right] \, ,
\eeq
with
\beq \label{en3}
I_n(Q_{\rm thre}) = \int_{Q_{\rm thre}}^\infty Q^{(n-1)/2} \bdRdQoFQ \, dQ \, .
\eeq
Physically, $\langle v \rangle$ is the average WIMP velocity, while $\langle
v^2 \rangle$ gives the velocity dispersion.\footnote{The dispersion of the
  function $f_1$ in the statistical sense is given by $\langle v^2 \rangle -
  \langle v \rangle^2$.} We emphasize that Eqs.(\ref{eqn215}) and (\ref{en3})
can be evaluated directly once the recoil spectrum is known; knowledge of the
functional form of $f_1(v)$ is not required.

Note that the first term in Eq.(\ref{eqn215}) vanishes for $n \geq 0$ if
$Q_{\rm thre} \rightarrow 0$. In the same limit, $\langle v^0 \rangle
\rightarrow \calN \alpha I_0(0)/2 \rightarrow 1$ by virtue of
Eq.(\ref{eqn213}). On the other hand, as written in Eqs.(\ref{eqn212}) and (\ref{eqn213}),
the velocity distribution integrated over the experimentally
accessible range of WIMP velocities gives a value smaller than unity. Using
only quantities that can be measured in the presence of a nonvanishing energy
threshold $Q_{\rm thre}$, we can replace Eq.(\ref{eqn213}) by
\beq \label{en4}
\calN(Q_{\rm thre}) = \frac {2}{\alpha} \left[ \frac
    {2 \~ Q_{\rm thre}^{1/2}} {F^2(Q_{\rm thre})} \afrac {dR} {dQ}_{Q=Q_{\rm thre}}
 + I_0(Q_{\rm thre}) \right]^{-1} \, .
\eeq
Using $\calN(Q_{\rm thre})$ in Eq.(\ref{eqn212}) ensures that the velocity
distribution integrated over $v \geq v_{\rm min}(Q_{\rm thre})$ gives unity.

We emphasize again that our final results in Eqs.(\ref{eqn212}) and
(\ref{eqn215}) are independent of the as yet unknown quantity $\calA$ defined
in Eq.(\ref{eqn202}). They do, however, depend on the WIMP mass $m_{\chi}$
through the coefficient $\alpha$ defined in Eq.(\ref{eqn205}).  This mass can
be extracted from a single recoil spectrum only if one makes some assumptions
about the velocity distribution $f_1(v)$. In the kind of model--independent
analysis pursued here, $m_{\chi}$ has to be determined by requiring that the
recoil spectra using two (or more) different target nuclei lead to the same
distribution $f_1(v)$ through Eq.(\ref{eqn212}). Note that this can be done
independent of the detailed particle physics model, which determine the value
of $\sigma_0$ for the two targets. However, one will need to know both form
factors, which strongly depend on whether spin--dependent or spin--independent
interactions dominate \cite{susydm}. Within a given particle physics model,
the best determination of $m_{\chi}$ should eventually come from experiments
at high--energy colliders.

\subsection{Simple model distributions}

The simplest semi--realistic model halo is a Maxwellian halo. The normalized
one--dimensional velocity distribution function in the rest frame of our
galaxy is then \cite{susydm}
\beq \label{eqn221}
f_{1,\Gau}(v) = \frac{4}{\sqrt{\pi}} \afrac{v^2}{v_0^3} e^{-v^2/v_0^2}\, ,
\eeq
where $v_0 \approx 220~{\rm km/s}$ is the orbital speed of the Sun around the
galactic center, which characterizes the velocity of all virialized objects in
the solar vicinity. On the other hand, when we take into account the orbital
motion of the solar system around the galaxy, as well as the orbit of the
Earth around the Sun, the distribution function must be modified to
\cite{susydm} 
\beq \label{eqn222}
f_{1,\sh}(v,v_e) = \frac{1}{\sqrt{\pi}} \afrac{v}{v_e v_0}
\bbigg{e^{-(v-v_e)^2/v_0^2}-e^{-(v+v_e)^2/v_0^2}}\, .
\eeq
Here
\beq \label{eqn223}
v_e(t) = v_0 \left[1.05+0.07 \cos\afrac{2 \pi (t-t_p)}{1~{\rm yr}}\right],
\eeq
where $t_p \simeq$ June 2nd is the date on which the velocity of the Earth
relative to the WIMP halo is maximal \cite{detac}. Eq.(\ref{eqn223}) includes
the effect of the rotation of the Earth around the Sun (second term), but does
not allow for the possibility that the halo itself might rotate.

Substituting these two expressions into Eq.(\ref{eqn201}), one easily obtains
the corresponding recoil energy spectra that WIMP direct detection would
measure \cite{susydm}:
\beq \label{eqn224a}
\adRdQ_\Gau = \calA \FQ \afrac{2}{\sqrt{\pi} v_0} e^{-\alpha^2 Q/v_0^2} \, ,
\eeq
and
\beq \label{eqn225a}
\adRdQ_\sh = \calA \FQ \afrac{1}{2 v_e} \bbigg{\erf{\T\afrac{\alpha
      \sqrt{Q}+v_e}{v_0}}-\erf{\T\afrac{\alpha \sqrt{Q}-v_e}{v_0}}}\, .
\eeq
Here $\erf(x)$ is the error function, defined as
\beqN
\erf(x) = \frac{2}{\sqrt{\pi}} \int_0^{x} e^{-t^2} dt.
\eeqN
Hence
\beq \label{eqn224b}
\frac{1}{\FQ} \adRdQ_\Gau \propto e^{-\alpha^2 Q/v_0^2}\, ,
\eeq
and
\beq \label{eqn225b}
\frac{1}{\FQ} \adRdQ_\sh \propto \erf{\T \afrac{\alpha
    \sqrt{Q}+v_e}{v_0}}-\erf{\T \afrac{\alpha \sqrt{Q}-v_e}{v_0}}\, .
\eeq
For future reference we note that the (unrealistic) ``reduced'' spectrum
(i.e., the recoil spectrum divided by the squared nuclear form factor) in
Eq.(\ref{eqn224b}) is exactly exponential; this remains approximately true for
the potentially quasi--realistic spectrum in Eq.(\ref{eqn225b}) as well.

In order to test our formulae, we calculated $\expv{v}$ and $\expv{v^2}$, first
from the normalized distribution functions in Eqs.(\ref{eqn221}) and
(\ref{eqn222}),
\cheqna
\beq \label{eqn226a}
\expv{v}_\Gau = \afrac{2}{\sqrt{\pi}} v_0,
\eeq
\cheqnb
\beq \label{eqn226b}
\expv{v^2}_\Gau = \afrac{3}{2} v_0^2,
\eeq
\cheqn
and
\cheqna
\beq \label{eqn227a}
\expv{v}_\sh = \afrac{v_0}{\sqrt{\pi}} e^{-v_e^2/v_0^2}
  +\left(\frac{v_0^2}{2 v_e}+v_e\right) \erf\afrac{v_e}{v_0},
\eeq
\cheqnb
\beq \label{eqn227b}
\expv{v^2}_\sh = \afrac{3}{2} v_0^2+v_e^2.
\eeq
\cheqn
Then we used the spectra of Eqs.(\ref{eqn224b}) and (\ref{eqn225b}) in
Eqs.(\ref{eqn212}), (\ref{eqn213}) and (\ref{eqn215}), taking $Q_{\rm thre} =
0$. They reproduced the normalized velocity distribution functions in
Eqs.(\ref{eqn221}) and (\ref{eqn222}), as well as the results in
Eqs.(\ref{eqn226a}) to (\ref{eqn227b}).

\section{Application to experiment}

In the previous section we have derived formulae for the normalized
one--dimensional velocity distribution function of WIMPs, $f_1(v)$, and for
its moments $\expv{v^n}$, from the recoil--energy spectrum, $dR/dQ$. In order
to use these expressions, one would need a functional form for $dR/dQ$. In
practice this might result from a fit to experimental data. Note that our
expression for $f_1(v)$ in Eq.(\ref{eqn212}) requires knowledge not only of
$dR/dQ$, but also of its derivative with respect to $Q$, i.e., we need to know
both the spectrum and its slope. This will complicate the error analysis
considerably, if $dR/dQ$ is the result of a fit.

In this section we therefore go one step further, and derive expressions that
allow to reconstruct $f_1(v)$ and its moments {\em directly from the data}.
The assumption we have to make is that the sample to be analyzed only contains
signal events, i.e., is free of background. This should be possible in
principle: The most copious backgrounds (from radioactive $\beta$ and $\gamma$
decays) can be discriminated on an event--by--event basis in many modern WIMP
detectors. The remaining background is dominated by elastic scattering of fast
neutrons. While these look just like signal events, this background can -- at
least in principle -- be made arbitrarily small by careful shielding and the
use of a muon veto system (to veto cosmic ray induced neutrons). Having a
sample of pure signal events, we can proceed with a complete statistical
analysis of the precision with which we can reconstruct $f_1(v)$ and its
moments.

In the absence of a true experimental sample of this kind, we had to resort to
Monte Carlo experiments. To that end we wrote an unweighted event generator.
To do so, we had to specify the form factor $F(Q)$ appearing in
Eq.(\ref{eqn201}); this is the topic of the first subsection. We then proceed
to analyze the reconstruction of $f_1(v)$ and its moments in two subsequent
subsections.

\subsection{The elastic form factor}

We start by presenting the two most commonly used parameterizations of the
squared nuclear form factor $\FQ$ appearing in Eq.(\ref{eqn201}). We focus on
spin--independent scattering, which usually dominates the event rate;
moreover, spin--dependent form factors are still only poorly understood.

The simplest form factor is the exponential one, first introduced by Ahlen
{\it et al.} \cite{FQa} and Freese {\it et al.}  \cite{detac}:
\beq \label{eqn301}\
F_{\rm ex}^2(Q) = e^{-Q/Q_0}\, ,
\eeq
where $Q$ is the recoil energy transferred from the incident WIMP to the target
nucleus,
\cheqna
\beq
Q_0 = \frac{1.5}{m_{\rm N} R_0^2}
\eeq
is the nuclear coherence energy and
\cheqnb
\beq
R_0 = \left[0.3+0.91 \afrac{m_{\rm N}}{\rm GeV}^{1/3}\right]~{\rm fm}
\eeq
\cheqn
is the radius of the nucleus. 

The exponential form factor implies that the radial density profile of the
nucleus has a Gaussian form. This Gaussian density profile is simple, but not
very realistic. Engel has therefore suggested a more accurate form factor
\cite{FQb}, inspired by the Woods--Saxon nuclear density profile,
\beq \label{eqn302}\
F_{\rm WS}^2(Q) = \bfrac{3 j_1(q R_1)}{q R_1}^2 e^{-(q s)^2}\, .
\eeq
Here $j_1(x)$ is a spherical Bessel function,
\beq
q = \sqrt{2 m_{\rm N} Q}
\eeq
is the transferred 3--momentum, and
\cheqna
\beq
R_1 = \sqrt{R_{A}^2-5 s^2}
\eeq
with
\cheqnb
\beq
R_{A} \simeq  1.2 \~ A^{1/3}~{\rm fm}\, , ~~~~~~~~~~~~~~~~ s \simeq 1~{\rm fm}\, .
\eeq
\cheqn
In our simulation we used the more accurate Woods--Saxon form factor in
Eq.(\ref{eqn302}).

\subsection{Reconstructing \boldmath$f_1(v)$}

Since we assume a detector without directional sensitivity, a single event is
uniquely characterized by the measured recoil energy $Q$. Existing experiments
such as CRESST \cite{CRESST} and CDMS \cite{CDMS} can determine the recoil
energy quite accurately. We will see shortly that the statistical errors on
the reconstructed velocity distribution $f_1(v)$ will be quite sizable even
for next--generation experiments, given existing bounds on the scattering
rate. We can therefore to good approximation ignore the error of $Q$ in our
analysis.

In the following we do not distinguish between the recoil spectrum $dR/dQ$ and
the actual differential counting rate $dN/dQ$. Since $dR/dQ$ is usually given
per unit detector weight and unit time, the two quantities differ by a
multiplicative constant. This constant cancels in Eq.(\ref{eqn212}), since it
will also appear in the normalization constant (\ref{eqn213}). 

We divide the total energy range into $B$ bins with central points $Q_n$ and
widths $b_n,~n = 1,~2,~\cdots,~B$. In each bin, $N_n$ events will be recorded.
Our simulated data set can therefore be described by
\beq \label{e3_0}
{\T Q_n-\frac{b_n}{2}} \le    \Qni \le    {\T Q_n+\frac{b_n}{2}}
,
   ~~~~~~~~~~~~~~~~ 
i =   1,~2,~\cdots,~N_n, ~~~ 
     n = 1,~2,~\cdots,~B.
\eeq
The standard estimate for $dR/dQ$ at $Q=Q_n$ is then\footnote{As usual in the
  physics literature, we use the same symbol for the estimator, i.e., the
  ``measurement'', of $dR/dQ$ and for the ideal recoil spectrum.}
\beq \label{e3_1}
r_n \equiv \afrac {dR} {dQ}_{Q = Q_n} = \frac {N_n} {b_n}, ~~~~~~~~~~~~~~~~~~~ 
 n = 1,~2,~\cdots,~B.
\eeq
The squared statistical error on $dR/dQ$ is accordingly
\beq \label{e3_2}
\sigma^2 (r_n) = \frac {N_n} {b_n^2} \, .
\eeq

As noted earlier, we also need the {\em slope} of the recoil spectrum to
reconstruct the velocity distribution; see Eq.(\ref{eqn212}). A rather crude
estimator of this slope is
\beq \label{e3_3}
s_{1,n} \equiv  \frac {d} {dQ} \afrac {dR}{dQ}_{Q = Q_n} = \frac {
  N_n(Q > Q_n) - N_n(Q < Q_n) } {(b_n/2)^2}\, ,
\eeq
where $N_n(Q > Q_n)$ and $N_n(Q<Q_n)$ are the numbers of events in bin $n$
which have measured recoil energy $Q$ larger and smaller than $Q_n$,
respectively. This estimator is rather crude, since it only uses the
information in which half of its bin a given event falls. 

It is clear intuitively that an estimator that makes use of the exact
$Q-$value of each event should be better. This can e.g. be obtained from the
average $Q-$value in a given bin, defined as
\beq \label{e3_4}
\Bar Q_n = \frac {1}{N_n} \sum_{i=1}^{N_n} Q_{n,i}  \, .
\eeq
Taylor--expanding $dR / dQ$ around $Q = Q_n$, keeping terms up to linear
order, gives
\beq \label{e3_5}
\afrac {dR} {dQ}_{Q \simeq Q_n} \simeq \afrac {dR} {dQ}_{Q=Q_n}
 + (Q - Q_n) \bbrac{\frac {d} {dQ} \afrac {dR} {dQ}_{Q=Q_n}}
 = r_n + (Q - Q_n) s_n \, .
\eeq
Eq.(\ref{e3_5}) allows to calculate $\Bar Q_n$ as  expectation value of $Q$ in
the $n-$th bin:
\beq \label{e3_6}
\Bar Q_n - Q_n = \frac { \int_{Q_n - b_n/2}^{Q_n + b_n/2} (Q - Q_n) [r_n + s_n
  (Q - Q_n)] \~ dQ } {\int_{Q_n - b_n/2}^{Q_n + b_n/2} [r_n + s_n (Q - Q_n)] \~ dQ }
= \frac {s_n b_n^2} {12 r_n} \, .
\eeq
An improved estimator of the slope is thus
\beq \label{e3_7}
s_{2,n} = \frac {12 r_n \left( \Bar Q_n - Q_n \right) } {b_n^2} \, .
\eeq

A simple calculation shows that the estimator (\ref{e3_7}) indeed has a
smaller statistical error than the one in Eq.(\ref{e3_3}). The definition
(\ref{e3_3}) immediately implies
\beq \label{e3_8}
\sigma^2(s_{1,n}) = \frac {N_n} {(b_n/2)^4} = \frac {16 r_n} {b_n^3}\, ,
\eeq
where we have used Eqs.(\ref{e3_1}) and (\ref{e3_2}). In order to calculate
the statistical error of the estimator $s_{2,n}$, we first have to compute
\beq \label{e3_9}
\overline{(Q-Q_n)^2}_n = \frac { \int_{Q_n - b_n/2}^{Q_n + b_n/2} (Q - Q_n)^2
  [r_n + s_n  (Q - Q_n)] \~ dQ } {\int_{Q_n - b_n/2}^{Q_n + b_n/2} [r_n + s_n (Q
  - Q_n)] \~ dQ } = \frac {b_n^2} {12} \, .
\eeq
Treating the number of events and the average $Q-$value in a given bin as
independent variables and using\footnote{Strictly speaking, the denominator
  should be $N_n - 1$.}
\beqN
\sigma^2\abrac{\Bar Q_n - Q_n}
 = \frac {1} {N_n} \left[ \overline{(Q-Q_n)^2}_n - \abrac{\Bar Q_n - Q_n}^2 \right] , \,
\eeqN
this yields
\beq \label{e3_10}
\sigma^2(s_{2,n}) = \left( \frac {s_{2,n}} {r_n} \right)^2 \sigma^2(r_n)
+ \left( \frac {12 r_n} {b_n^2} \right)^2 \sigma^2\abrac{\Bar Q_n - Q_n}
= \frac {12 r_n} {b_n^3}\, .
\eeq
This is smaller than the error (\ref{e3_8}), by a factor 3/4.

An important observation is that the statistical error of both estimators of
the slope of the recoil spectrum scale like the bin width to the power
$-1.5$. This can intuitively be understood from the argument that the
variation of $dR/dQ$, which we are trying to determine, will be larger for
larger bins. One would therefore naively conclude that the errors of the
estimated slopes would be minimized by choosing very large bins.

However, both Eq.(\ref{e3_3}) and Eq.(\ref{e3_7}) reproduce the actual slope
at $Q = Q_n$ only if the Taylor expansion (\ref{e3_5}) holds; in the
presence of higher powers of $Q-Q_n$ neither of these estimates exactly
reproduces the true slope at $Q = Q_n$. Not surprisingly, the influence of
these higher powers, which induce uncontrolled systematic errors, will {\em
  increase} quickly with increasing bin width $b_n$.

\begin{figure}[t]
\begin{center}
\rotatebox{270}{\includegraphics[width=13cm]{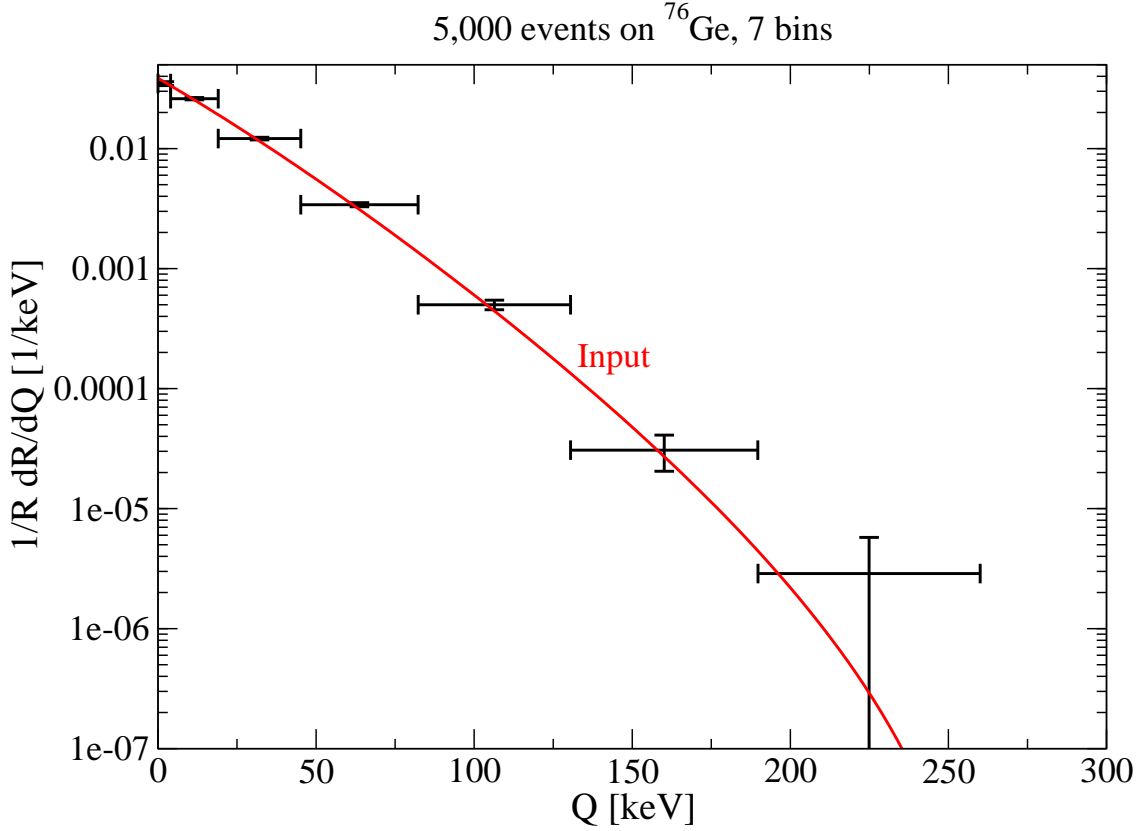}}
\end{center}
\caption{The curve shows the theoretical predicted recoil energy spectrum
  $(dR/dQ)_\sh$ for a shifted Maxwellian WIMP velocity distribution,
  and the Woods--Saxon form factor. The data points with error bars show
  simulated experimental data produced from this spectrum (5,000 total events,
  $m_{\chi}$=100 GeV/$c^2$, $m_{\rm N}$=70.6 GeV/$c^2$ for $\rmXA{Ge}{76}$,
  $v_0$=220 km/s, $v_e$=231 km/s), and galactic escape velocity $v_{\rm esc} = 700$
  km/s. The vertical error bars show the statistical uncertainties of the
  measurements, while the horizontal error bars indicate the bin widths.} 
\label{fig301}
\end{figure}

We had seen at the end of Sec.~2 that the predicted recoil spectrum resembles
a falling exponential. This is confirmed in Fig.~\ref{fig301}, which shows the predicted
recoil spectrum of a 100 GeV WIMP on $^{76}$Ge, using the Woods--Saxon
form factor (\ref{eqn302}) and the ``shifted Maxwellian'' velocity
distribution of Eq.(\ref{eqn222}); as usual, we cut the velocity distribution
off at a velocity $v_{\rm esc}$, here taken to be 700 km/s, since WIMPs with $v > v_{\rm esc}$
can escape the gravitational well of our galaxy. This figure also shows the
result of a simulated experiment, where the exposure time and cross section
are chosen such that the expected number of events is 5,000; these have been
collected in seven bins in recoil energy.

While an approximately exponential function can be approximated by a linear
ansatz, as in Eq.(\ref{e3_5}), only over a narrow range of $Q$, i.e., for small
bin widths, the {\em logarithm} of this function can be approximated by a
linear ansatz for much wider bins. This corresponds to the ansatz
\beq \label{eqn311}
 \adRdQ_n \equiv
 \afrac {dR} {dQ}_{Q \simeq Q_n} \simeq \Td{r}_n \~ {\rm e}^{\kn (Q-Q_n)}
 \equiv \rn \~ {\rm e}^{\kn (Q - Q_{s,n})}\, .
\eeq
Here $\Td{r}_n$ is the recoil spectrum at the point $Q=Q_n$,
\beq
\Td{r}_n\equiv \adRdQ_{Q=Q_n}\, ,
\eeq
while $k_n$ is the {\em logarithmic slope} of the recoil spectrum in the
$n-$th bin. 

Our next task is to find estimators for $\Td{r}_n$ and $k_n$ using
(simulated) data. Note that for $\kn \neq 0$, $\Td{r}_n$ cannot be
estimated from the number of events $N_n$ in the $n-$th bin alone. Instead,
one has
\beq \label{eqn312}
N_n = \intQnbn \adRdQ_n dQ = b_n \Td{r}_n\afrac{\sinh x_n} {x_n}\, , 
\eeq
where we have introduced the dimensionless quantities
\beq \label{e3_11}
x_n \equiv \frac {b_n k_n} {2} \, .
\eeq
Hence,
\beq
\Td{r}_n= \frac{N_n}{b_n} \afrac{x_n}{\sinh x_n}
\label{eqn313}
\eeq
depends on $k_n$.
On the other hand, the second, equivalent expression in Eq.(\ref{eqn311})
still uses the quantities $r_n = N_n/b_n$ as normalization. Evidently, they
describe the spectrum $dR/dQ$ at the shifted points $Q_{s,n}$. Equivalence of
the two expressions in Eq.(\ref{eqn311}) implies
\beq \label{e3_12}
Q_{s,n} = Q_n + \frac{1}{k_n} \ln \left( \frac {\sinh x_n} {x_n} \right) \, .
\eeq
The second expression in Eq.(\ref{eqn311}) thus has the advantage that the
prefactor $r_n$ can be computed more easily than $\Td{r}_n$; on the
other hand, while the $Q_n$ are simply the midpoints of the $n-$th bin, and
can thus be chosen at will, the $Q_{s,n}$ are derived quantities; they depend
on the logarithmic slopes $k_n$, which we haven't determined yet.

To do so, we again use the average $Q-$value in the $n-$th bin. From
Eq.(\ref{eqn311}) we find:
\beq \label{eqn314}
\Bar Q_n - Q_n = \frac {\intQnbn(Q-Q_n) \adRdQ_n dQ}
{\intQnbn \adRdQ_n dQ} = \afrac{b_n}{2}  \coth x_n  - \frac{1}{k_n} \, . 
\eeq
Unfortunately this expression cannot be solved analytically for $k_n$. It is,
however, straightforward to find $k_n$ numerically once $\Bar Q_n$ is
known. Alternatively, we can make use of the second moment of the recoil
spectrum in the $n-$th bin, defined as
\beq \label{eqn315}
\overline{(Q-Q_n)^2}_n = \frac {\intQnbn (Q-Q_n)^2 \adRdQ_n dQ} {\intQnbn \adRdQ_n dQ} =
\afrac{b_n}{2}^2 \left[ 1 - 2 \afrac{\coth x_n}{x_n} + 
\frac {2} {x_n^2} \right] \,.
\eeq
Multiplying both sides of Eq.(\ref{eqn314}) with $b_n/x_n$ and adding to
Eq.(\ref{eqn315}), we can calculate the logarithmic slopes as
\beq \label{eqn316}
\kn = \frac{8 \left( \Bar Q_n - Q_n \right) } {b_n^2-4 \~ \overline{(Q-Q_n)^2}_n} \, .
\eeq
Note that $\kn$ determined either from Eq.(\ref{eqn314}) or from
Eq.(\ref{eqn316}) is independent of the normalization $r_n$ or
$\Td{r}_n$. 

In the following we will estimate the logarithmic slopes from
Eq.(\ref{eqn314}), since it simplifies the error analysis somewhat; note that
the statistical errors of $\Bar Q_n$ and $\overline{(Q-Q_n)^2}_n$ are correlated. Using
standard error propagation, we have
\beq \label{e3_13}
\sigma^2(k_n) = \sigma^2\abrac{\Bar Q_n - Q_n} \cdot
 \left[ \frac { d \abrac{\Bar Q_n - Q_n}} {d k_n} \right]^{-2}\, .
\eeq
\vspace*{-6.5mm}
\begin{figure}[h!]
\begin{center}
\rotatebox{270}{\includegraphics[width=10cm]{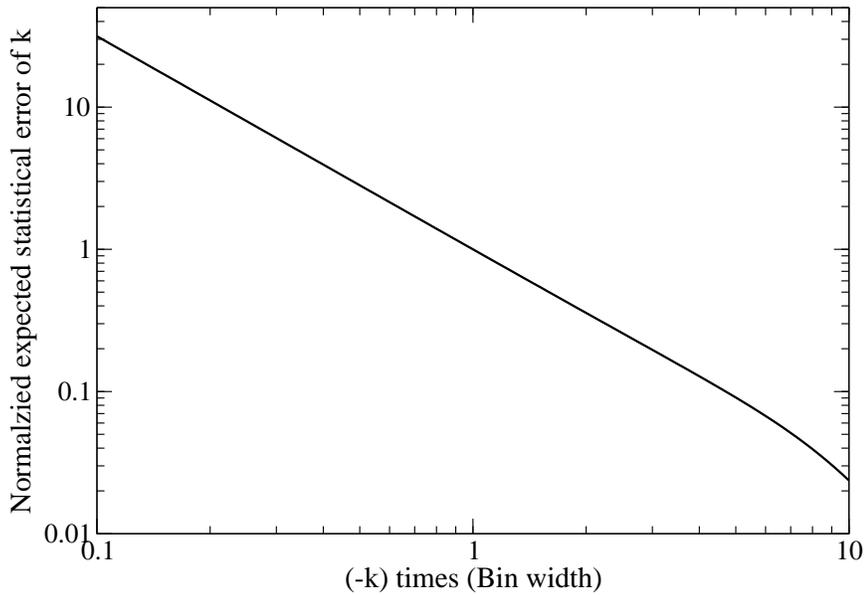}}
\end{center}
\caption{The expected statistical error of our estimator for the logarithmic
  slope based on Eq.(\ref{eqn314}) as function of the bin width. The bin width
  is given in units of the inverse logarithmic slope, and the error is
  normalized to its value for $b_n = -1/k_n$. In this calculation we have
  assumed that the ansatz (\ref{eqn311}) is exact within the given bin. }
\label{fig302}
\end{figure}
The error on the average energy transfer can be estimated directly from the
data, using
\beq \label{e3_14}
\sigma^2\abrac{\Bar Q_n - Q_n}
 = \frac {1} {N_n - 1} \left[ \overline{(Q-Q_n)^2}_n - \abrac{\Bar Q_n - Q_n}^2 \right] , \,
\eeq
where now $\overline{(Q-Q_n)^2}_n$ is estimated from the data, analogously to the experimental
definition of $\Bar Q_n$ in Eq.(\ref{e3_4}). The second factor in
Eq.(\ref{e3_13}) can be calculated straightforwardly from Eq.(\ref{eqn314}):
\beq \label{e3_15}
\left[ \frac { d \abrac{\Bar Q_n - Q_n}} {d k_n} \right]^{-1} =  \frac {k_n^2}
{f(x_n)} \, ,
\eeq
where we have defined the auxiliary function
\beq \label{e3_15a}
f(x) = 1 - \left( \frac {x} {\sinh x} \right)^2\, .
\eeq

For given input values $r_n, \, k_n$ and $b_n$, Eqs.(\ref{e3_13}) and
(\ref{e3_15}) also allow to calculate the expected statistical error of the
estimated $k_n$, using Eq.(\ref{eqn315}) to calculate the expected error of
$\Bar Q_n - Q_n$. The result is shown in Fig.~\ref{fig302}. By normalizing the bin width
to the inverse slope, and the expected error to its value for a given bin
width, the result becomes independent of $r_n$, and can in fact be used for
all combinations of $k_n$ and $b_n$. We observe that for small bins, the
expected error again scales like $b_n^{-1.5}$, just as the expected errors
(\ref{e3_8}) and (\ref{e3_10}) of our two estimators for the linear slope. If
the bin width is significantly larger than the absolute value of the inverse
of the logarithmic slope, the error decreases even faster with increasing bin
width.\footnote{Given the exponential form of our ansatz (\ref{eqn311}) one
  might assume that the statistical error of the estimated values of the
  $k_n$ could be minimized by estimating them from the average values of
  $\exp[\kappa(Q-Q_n)]$, for some fixed value of $\kappa$. For sufficiently
  small $|\kappa|$ this in fact amounts to using the average value of $Q$, as
  described in the text. Increasing $|\kappa|$ leads to slightly {\em larger}
  expected statistical errors.}

This again argues in favor of using large bins. However, we again have to
consider systematic errors. After all, it is quite unlikely that the (as yet
unknown) recoil spectrum $dR / dQ$ exactly satisfies our ansatz (\ref{eqn311})
over an extended range of $Q$. Rather, this ansatz should be considered as the
first terms in a Taylor expansion of the logarithm of $dR / dQ$. In this case
the next--order term, which contributes $c_n (Q - Q_n)^2$ in the exponent,
will already modify $\Bar Q_n$. Since we estimate $k_n$ from the numerical
value of $\Bar Q_n$ using Eq.(\ref{eqn314}), which is exact only for $c_n =
0$, any non--zero $c_n$ will introduce some systematic error in our estimate
of $k_n$.

Fortunately much of this error can be absorbed by a simple trick. According to
(\ref{eqn311}) the logarithmic slope is constant over the entire bin, i.e., we
could use $k_n$ extracted from Eq.(\ref{eqn314}) as estimate of the
logarithmic slope at any point $Q$ between $Q_n - b_n/2$ and $Q_n + b_n/2$.
Once $c_n \neq 0$ the true logarithmic slope will in fact vary with $Q$.
However, one may hope that the expectation value of our estimator still
reproduces the true logarithmic slope at {\em some} value of $Q$ within the
$n-$th bin.

\begin{figure}[t!]
\begin{center}
\rotatebox{270}{\includegraphics[width=12.5cm]{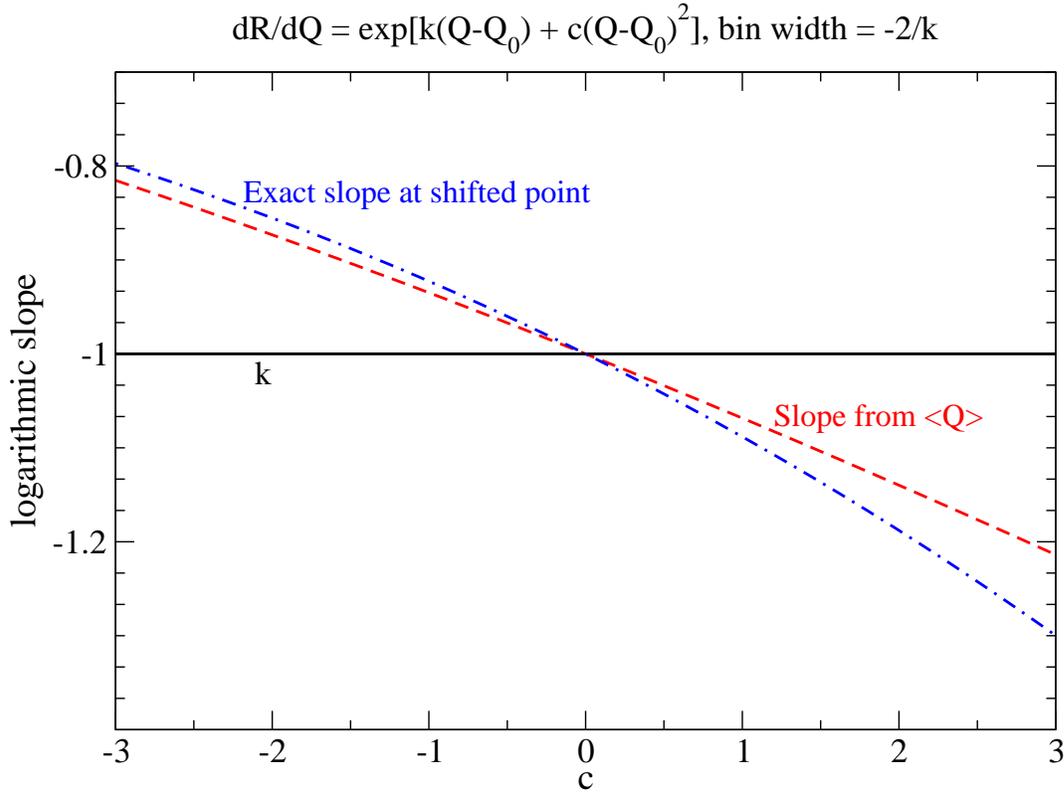}}
\end{center}
\caption{Illustration of the systematic error of our estimator for the
  logarithmic slope based on Eq.(\ref{eqn314}). Here we assume that $dR/dQ
  \propto \exp[k(Q-Q_0) + c(Q-Q_0)^2]$, i.e., we amend our ansatz
  (\ref{eqn311}) by adding a quadratic term to the exponent. The horizontal
  black line shows the input value of $k$, which now describes the true
  logarithmic slope only at $Q = Q_0$. The dashed (red) line shows the
  expectation value of our estimator for the logarithmic slope, while the
  dot--dashed (blue) line shows the true logarithmic slope at the shifted
  point $Q_s$ defined in Eq.(\ref{e3_12}). The result holds for $k = -1$ and a
  bin width $b = 2$.} 
\label{fig303}
\end{figure}

This is illustrated by Fig.~\ref{fig303}, which shows various evaluations of
the logarithmic slope within one bin as function of the quadratic coefficient
$c$. The true logarithmic slope at the center of the bin is, of course, still
given by $k$, independent of the correction $c$. As argued above, the
expectation value of our estimator, shown by the dashed (red) line, does
depend on $c$. Note, however, that our estimator comes quite close to the true
logarithmic slope at the shifted value $Q_s$ defined in Eq.(\ref{e3_12}),
which is shown by the dot--dashed (blue) line; this is true for both signs of
$c$. We therefore conclude that we can minimize the leading systematic error
by interpreting our estimator of $k_n$ as logarithmic slope of the recoil
spectrum, not at the center of the bin $Q_n$, but at the shifted point
$Q_{s,n}$. Note that $Q_{s,n}$ itself depends on $k_n$; this, however, does
not introduce any additional error, if we simply interpret Eq.(\ref{e3_12}) as
an -- admittedly somewhat complicated -- prescription for the determination of
the $Q-$values where we wish to estimate the logarithmic slope of the recoil
spectrum.

Using large bins has a second, obvious disadvantage: the number of bins scales
inversely with their size, i.e., by using large bins we'd be able to estimate
$f_1$ only at a small number of velocities. This can be alleviated by using
overlapping bins, or -- equivalently -- by combining several relatively small
bins into overlapping ``windows''. This means that a given data point
$Q_{n,i}$ may well contribute to several different windows, and hence to the
measurement of $f_1$ at several values of $v$. This can increase the total
amount of information about $f_1$ since the only information we use about the
data points in a given window is encoded in the average recoil energy in this
window. This averaging effectively destroys information. By letting a given
data point contribute to several overlapping windows, this loss of information
can be reduced.

A final disadvantage of using large bins or windows is that it would lead to a
quite large minimal value of $v$ where $f_1$ can be reconstructed, simply
because the central value $Q_1$, and also the shifted point $Q_{s,1}$, of a
large first bin would be quite large. This can be again be alleviated by first
collecting our data in relatively small bins, and then combining varying
numbers of bins into overlapping windows. In particular, the first window
would be identical with the first bin.

A final consideration concerns the size of the bins. Choosing fixed bin sizes,
and therefore also mostly fixed window sizes, would lead to errors on the
estimated logarithmic slopes, and hence also on the estimates of $f_1$, that
increase quickly with increasing $Q$ or $v$. This is due to the essentially
exponential form of the recoil spectrum, which would lead to a quickly falling
number of events in equal--sized bins. We found that we get roughly equal
errors in all bins if we instead take linearly increasing bins.

These considerations motivate the following set--up for our mock experimental
analysis. We start by binning the data, as in Eq.(\ref{e3_0}), where the bin
widths satisfy
\beq \label{e3_16}
b_n = b_1 + (n-1) \delta \, ;
\eeq
here the increment $\delta$ satisfies 
\beq \label{e3_17}
\delta = \frac{2}{B(B-1)} \aBig { Q_{\rm max} - Q_{\rm min} - B b_1} \, ,
\eeq
$B$ being the total number of bins, and $Q_{\rm max, min}$ being the
(kinematical or instrumental) extrema of the recoil energy. We then collect up
to $n_W$ bins into a window, with smaller windows at the borders of the range
of $Q$. In the following we use Latin indices $i,~j,~\cdots$ to label bins, and
Greek indices $\mu,~\nu,~\cdots$ to label windows; later on we will use Latin
indices $a,~b,~\cdots$ to label all events in the sample. For $1 \leq \mu \leq
n_W$ the $\mu-$th window simply consists of the first $\mu$ bins; for $n_W
\leq \mu \leq B$, the $\mu-$th window consists of bins $\mu-n_W+1,~\mu-n_W+2,
\cdots,~\mu$; and for $B \leq \mu \leq B + n_W - 1$, the $\mu-$th window
consists of last $n_W - \mu + B$ bins. This can also be described by
introducing the indices $i_{\mu-}$ and $i_{\mu+}$ which label the first and
last bin contributing to the $\mu-$th window, with
\beq \label{e3_18}
i_{\mu-} = \left\{ \begin{array}{ll} 1, & \mu \leq n_W \\
\mu - n_W + 1, & \mu \geq n_W \end{array}\right., \ \ \ \
i_{\mu+} = \left\{ \begin{array}{ll} \mu, & \mu \leq B \\
B, & \mu \geq B \end{array}\right., \ \ \ (1 \leq \mu \leq B +
n_W - 1) \, .
\eeq
The center of the $i-$th bin is called $Q_i$, as before. The total number of
windows defined through Eq.(\ref{e3_18}) is evidently $W = B + n_W - 1$.

The basic observables needed for the reconstruction of $f_1$ are then the
number of events $N_i$ in the $i-$th bin, as well as the average $\Bar Q_i$
defined as in Eq.(\ref{e3_4}). From these one easily
calculates the number of events per window,
\beq \label{e3_19}
N_\mu = \sum_{i = i_{\mu-}}^{i_{\mu+}} N_i \,
\eeq
as well as the averages
\beq \label{e3_20}
\Bar Q_\mu = \frac {1}{N_\mu} \sum_{i = i_{\mu-}}^{i_{\mu+}} N_i \Bar Q_i \,. 
\eeq

One drawback of the use of overlapping windows in the analysis is that the
observables defined in Eqs.(\ref{e3_19}) and (\ref{e3_20}) are all correlated
(for $n_W \neq 1$). The slope in a given window will again be calculated as in
Eq.(\ref{eqn314}), with ``bin'' quantities replaced by ``window''
quantities. We thus need the covariance matrix for the $\Bar Q_\mu -
Q_\mu$, where $Q_\mu$ is the midpoint of the $\mu-$th window; it follows
directly from the definition (\ref{e3_20}):
\beq \label{e3_21}
{\rm cov}\left(\Bar Q_\mu - Q_\mu, \Bar Q_\nu - Q_\nu\right) = \frac
{1}{N_\mu N_\nu} \sum_{i = i_{\nu-}}^{i_{\mu+}} \bigg[ N_i^2
  \sigma^2\left(\Bar Q_i - Q_i\right) + N_i \left( \Bar Q_i - \Bar Q_\mu \right)
\left( \Bar Q_i - \Bar Q_\nu \right) \bigg] \, ,
\eeq
where $\sigma^2(\Bar Q_i - Q_i)$ is defined as in Eq.(\ref{e3_14}). In
Eq.(\ref{e3_21}) we have assumed $\mu \leq \nu$; the covariance matrix
is, of course, symmetric. Moreover, the sum is understood to vanish if the two
windows $\mu, \ \nu$ do not overlap, i.e., if $i_{\mu+} < i_{\nu-}$.

The ansatz (\ref{eqn311}) is now assumed to hold over an entire window. We
again estimate the prefactor as
\beq \label{e3_22}
r_\mu = \frac {N_\mu} {w_\mu}\, ,
\eeq
$w_\mu$ being the width of the $\mu-$th window. This implies
\beq \label{e3_23}
{\rm cov}(r_\mu, r_\nu) = \frac {1} {w_\mu w_\nu}  \sum_{i =
  i_{\nu-}}^{i_{\mu+}} N_i \, ,
\eeq
where we have again taken $\mu \leq \nu$. Finally, the mixed covariance
matrix is given by
\beq \label{e3_24}
{\rm cov}\left(r_\mu, \Bar Q_\nu - Q_\nu\right) = \frac {1} {w_\mu N_\nu}
\sum_{i = i_-}^{i_+} N_i \left( \Bar Q_i - \Bar Q_\nu \right) \, .
\eeq
This sub--matrix is not symmetric under the exchange of $\mu$ and
$\nu$. In the definition of $i_-$ and $i_+$ we therefore have to distinguish
two cases:
\begin{eqnarray} \label{e3_25}
{\rm If} \ \mu \leq \nu\&:\& \ i_- = i_{\nu-}\,, \ i_+ = i_{\mu+} \, ;
\nonumber \\
{\rm If} \ \mu \geq \nu\&:\& \ i_- = i_{\mu-}\,, \ i_+ = i_{\nu+} \, .
\end{eqnarray}
As before, the sum in Eq.(\ref{e3_24}) is understood to vanish if $i_- > i_+$.

The covariance matrices involving our estimators of the logarithmic slopes
$k_\mu$, derived from Eq.(\ref{eqn314}) with $n \rightarrow \mu$
everywhere, can be calculated in terms of the covariance matrices in
Eqs.(\ref{e3_21}) and (\ref{e3_24}):
\beq \label{e3_26}
{\rm cov}(k_\mu, k_\nu) = \bfrac { k_\mu^2 k_\nu^2} { f(x_\mu)
  f(x_\nu) } {\rm cov}\left(\Bar Q_\mu - Q_\mu, \Bar Q_\nu - Q_\nu\right)\, ,
\eeq
where $x_\mu$ is as in Eq.(\ref{e3_11}) with $n \rightarrow \mu$, and the
function $f(x)$ has been defined in Eq.(\ref{e3_15a}); and
\beq \label{e3_26a}
{\rm cov}(r_\mu, k_\nu) = \bfrac {k_\nu^2} {f(x_\nu)} {\rm cov} \left(r_\mu,
\Bar{Q}_\nu - Q_\nu\right) \, .
\eeq

We are now ready to put all pieces together to compute the reconstructed
velocity distribution and its statistical error. Inserting the ansatz
(\ref{eqn311}) with the substitution $n \rightarrow \mu$ into
Eq.(\ref{eqn212}), one finds the reconstructed velocity distribution
\beq \label{e3_27}
f_{1,r}(v_\mu) = {\cal N}  \bfrac{2 Q_{s,\mu} r_\mu }{F^2(Q_{s,\mu})}
\bbrac{ \frac {d} {dQ} \ln F^2(Q) \bigg|_{Q = Q_{s,\mu}} - k_\mu} \, .
\eeq
Here, $Q_{s,\mu}$ is given by Eq.(\ref{e3_12}) with $n \rightarrow \mu$,
and
\beq \label{e3_28}
v_\mu = \alpha \sqrt{Q_{s,\mu}}, \,
\eeq
see Eq.(\ref{eqn204}). Finally, the normalization ${\cal N}$ defined in
Eq.(\ref{eqn213}) can be estimated directly from the data:
\beq \label{e3_29}
{\cal N}^{-1} = \frac {\alpha} {2} \sum_a \frac {1} { \sqrt{Q_a} F^2(Q_a) }\,
\
\eeq
where the sum runs over all events in the sample. 

Since neighboring windows overlap, the estimates of $f_1$ at adjacent values
of $v_\mu$ are correlated. This is described by the covariance matrix
\begin{eqnarray} \label{e3_30}
 \conti {\rm cov} \aBig{ f_{1,r}(v_\mu), f_{1,r}(v_\nu)} \non\\
 \= 
\bfrac { f_{1,r}(v_\mu) f_{1,r}(v_\nu) } { r_\mu r_\nu } {\rm cov} (r_\mu,
r_\nu)
+ ( 2 {\cal N} )^2 \bfrac {Q_{s,\mu} Q_{s,\nu} r_\mu  r_\nu} { F^2(Q_{s,\mu})
  F^2(Q_{s,\nu}) } {\rm cov}(k_\mu, k_\nu)
\nonumber \\
\conti ~~~~~~
 - 2 {\cal N} \cbrac{ \frac { Q_{s,\mu} r_\mu } {F^2(Q_{s,\mu})} 
\bfrac {f_{1,r}(v_\nu)} {r_\nu} {\rm cov}(k_\mu, r_\nu)
+ (\mu \longleftrightarrow \nu)}\, .
\end{eqnarray}
The covariance matrices involving the normalized counting rates $r_\mu$ and
logarithmic slopes $k_\mu$ have been given in Eqs.(\ref{e3_23}), (\ref{e3_26})
and (\ref{e3_26a}). In principle Eq.(\ref{e3_30}) should also include
contributions involving the statistical error of our estimator (\ref{e3_29})
for ${\cal N}$. However, we find this error, and its correlations with the
errors of the $r_\mu$ and $k_\mu$, to be negligible compared to the errors
included in Eq.(\ref{e3_30}). 

We are finally in a position to present some numerical results. We first
validate our results by presenting $\chi^2_f$ distributions, defined via
\beq \label{e3_31} 
\chi^2_f \equiv \frac {1} {W} \sum_{\mu,\nu} {\cal C}_{\mu\nu}
\left[ f_{1,r}(v_\mu) - f_1(v_\mu) \right] \left[ f_{1,r}(v_\nu) - f_1(v_\nu)
\right] \, .  
\eeq
Here $f_{1,r}$ is our estimate (\ref{e3_27}) of the velocity distribution,
$f_1$ is the true (input) distribution, and ${\cal C}$ is the inverse of the
covariance matrix of Eq.(\ref{e3_30}). We expect $\chi_f^2$ to be (roughly)
distributed according to the standard $\chi^2$ distribution when the results
of sufficiently many simulated experiments, with sufficiently many events per
experiment, are analyzed.

\begin{figure}[h!]
\begin{center}
\rotatebox{270}{\includegraphics[width=10.5cm]{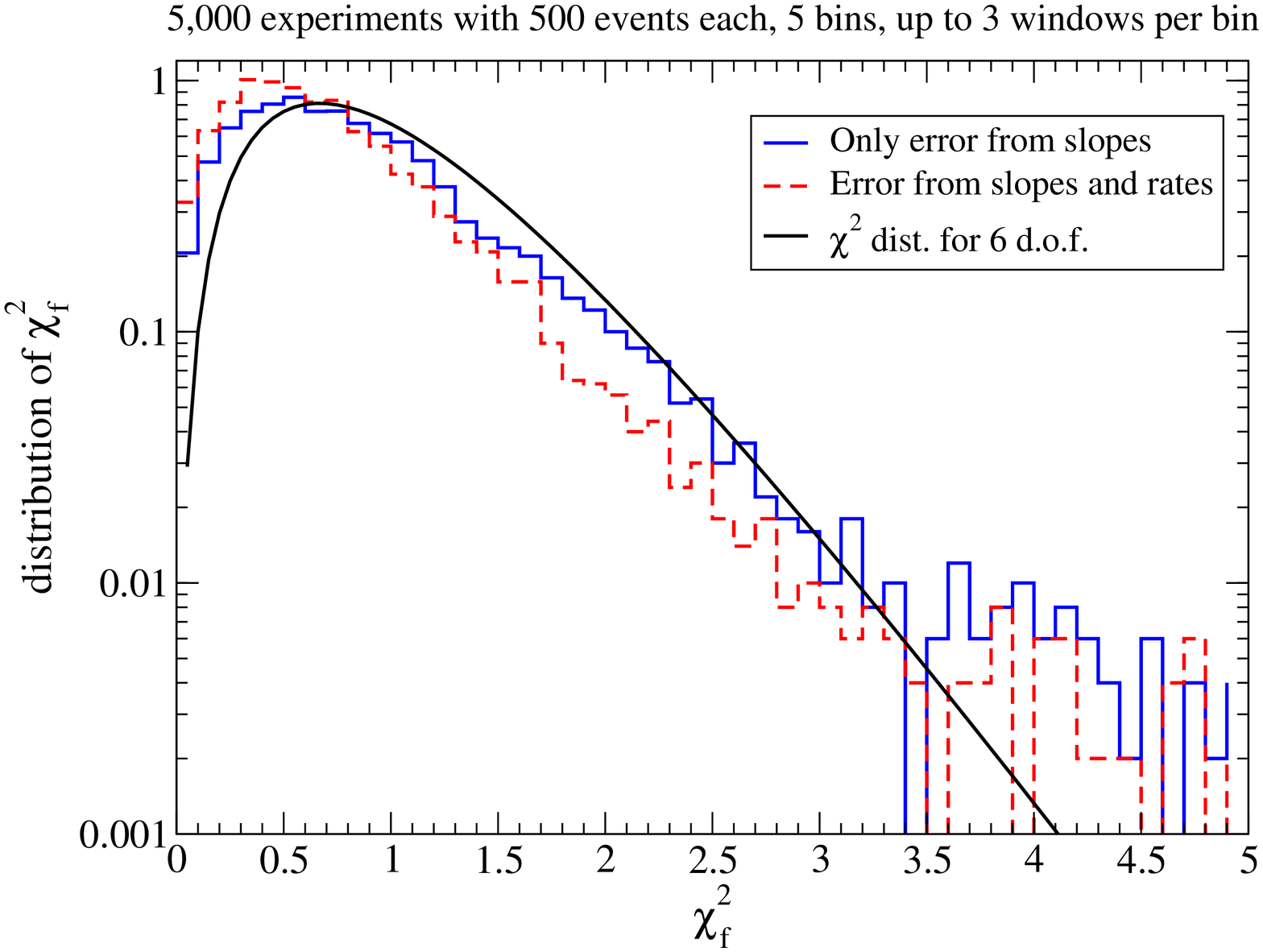}} \\
\vspace*{-5mm}
\rotatebox{270}{\includegraphics[width=10.5cm]{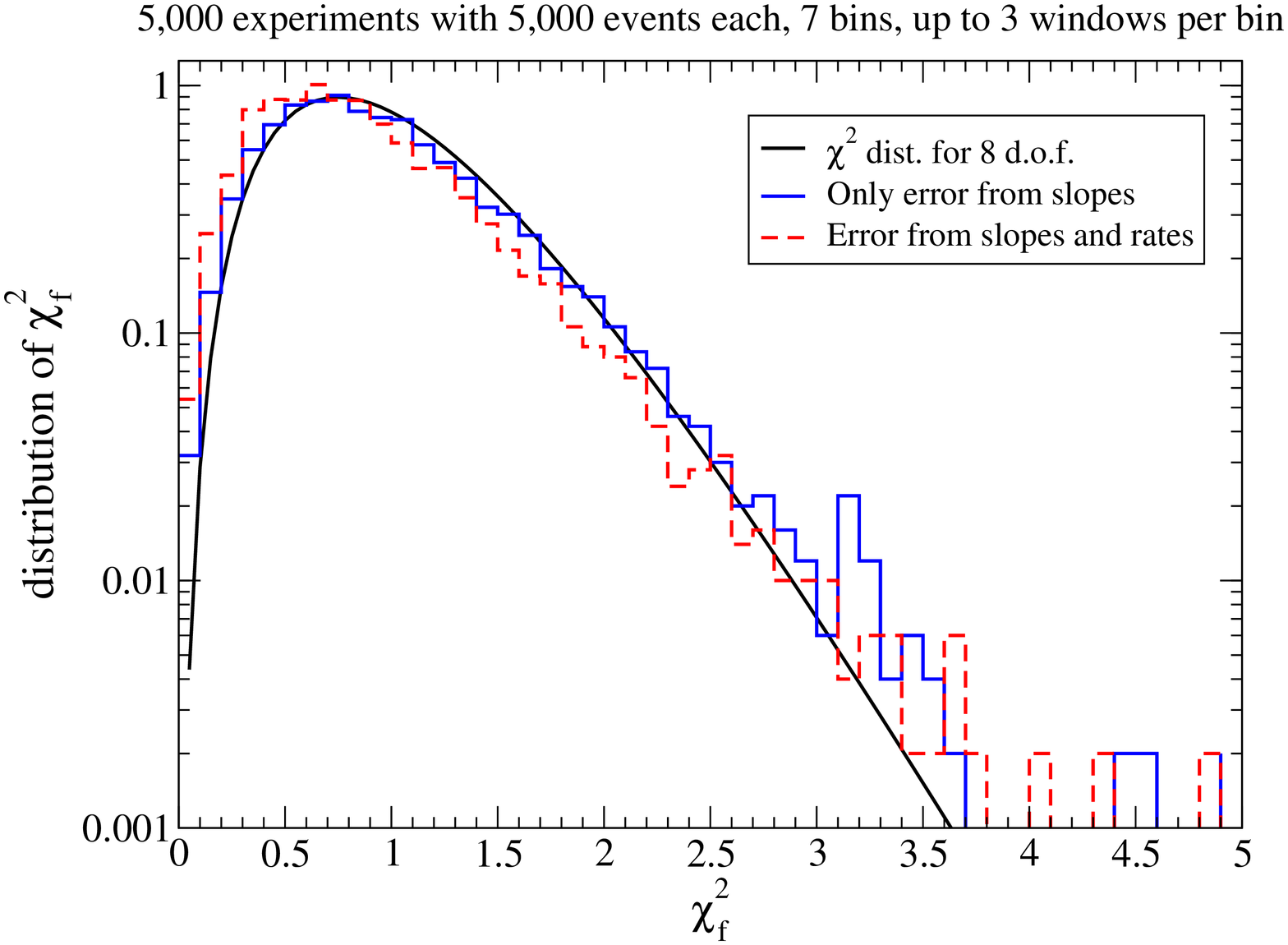}}
\end{center}
\vspace*{-5mm}
\caption{The distribution of $\chi^2_f$ defined in Eq.(\ref{e3_31}) over 5,000
  simulated experiments, with on average 500 (top) and 5,000 (bottom) events per
  experiment. The solid (blue) and dashed (red) histograms have been obtained
  by estimating the covariance matrix for the reconstructed velocity
  distribution excluding (including) the statistical errors on the number of
  events in the windows. The smooth (black) curves show the theoretical
  $\chi^2$ distributions for the appropriate numbers of degrees of
  freedom. Parameters are as in Fig.~\ref{fig301}.}
\label{fig304}
\end{figure}

Figs.~\ref{fig304} show $\chi^2_f$ distributions for 5,000 simulated
experiments, with on average 500 (top) and 5,000 (bottom) events per
experiments. Note that the actual number of events in a given simulated
experiment varies according to the Poisson distribution; otherwise one would
introduce an artificial correlation between the normalized counting rates
$r_i$ in different bins. Moreover, the number of bins has been fixed a priori
in these analyses. The last bin is typically empty, and has therefore been
ignored in the analysis. This also reduces the number of windows used in the
analysis by one, i.e., the upper (lower) frame shows results for $W=6$ ($W=8$).

The two histograms in each figure differ by the number of terms that have been
included in the estimate of the covariance matrix for $f_{1,r}$. The solid
(blue) histograms have been obtained by only including the second term in
Eq.(\ref{e3_30}), while the dashed (red) histograms also include the other
terms, which are due to the statistical errors on the rescaled event numbers
$r_\mu$. We note that including these terms on average leads to a slight
overestimate of the true error of $f_{1,r}$, i.e., the average of $\chi_f^2$ is
somewhat smaller than unity. This is partly due to the fact that we have
ignored the error on the normalization ${\cal N}$, which is correlated quite
strongly with the errors on the $r_\mu$.

The lower figure demonstrates that for an average of 5,000 events per
experiment the distribution of $\chi_f^2$ values becomes quite similar to the
well--known $\chi^2$ distribution, shown by the smooth curve. At least two
effects contribute to the difference. First, we heavily relied on Gaussian
error propagation in our estimate (\ref{e3_30}) of the covariance matrix of
the reconstructed velocity distribution. This is essentially a Taylor
expansion, including only the first non--trivial term. It therefore becomes
exact only in the limit of small errors, i.e., for large numbers of events in a
given window. Since the recoil spectrum is falling essentially exponentially,
this condition is practically always violated at least in the last bin(s), see
Fig.~\ref{fig301}. We discard windows containing less than 3 events, but it is
clear that this at best alleviates the problem. In the case at hand, this
evidently results in an overestimate of the true error. Secondly, our estimator
(\ref{e3_27}) for the velocity distribution relies on the estimate of the
logarithmic slopes $k_\mu$, which in turn is based on Eq.(\ref{eqn314}). As
illustrated in Fig.~\ref{fig303} this estimate of $k_\mu$ in general has some
systematic error, which would tend to increase $\chi_f^2$. However, this
figure also led us to expect small systematic errors if $k_\mu$ is interpreted
as estimator of the logarithmic slope at the shifted points $Q_{s,\mu}$.
Indeed, as stated above, the total expression (\ref{e3_30}) somewhat
overestimates the true error even in the lower frame of Figs.~\ref{fig304},
which assumes a large number of events but uses a rather small number of bins,
which thus have to be quite large. Had we instead interpreted $k_\mu$ as
estimator of the slope at $Q_\mu$, the average $\chi_f^2$ would be about 2.9,
indicating that the systematic error would have dominated.

Figs.~\ref{fig304} also show an excess of simulated experiments with rather
large values of $\chi_f^2$ if the covariance matrix for $f_{1,r}$ is estimated
based on the errors on the $k_\mu$ only. This is true also for the upper
frame, even though in this case the average value of $\chi_f^2$ is only about
0.93. To be conservative, from now on we therefore take the full
Eq.(\ref{e3_30}) as our estimator of the covariance matrix of $f_{1,r}$,
leading to average $\chi_f^2 = 0.78\ (0.94)$ for the upper (lower) frame of
Figs.~\ref{fig304}. 

\begin{figure}[t!]
\begin{center}
\rotatebox{270}{\includegraphics[width=10.5cm]{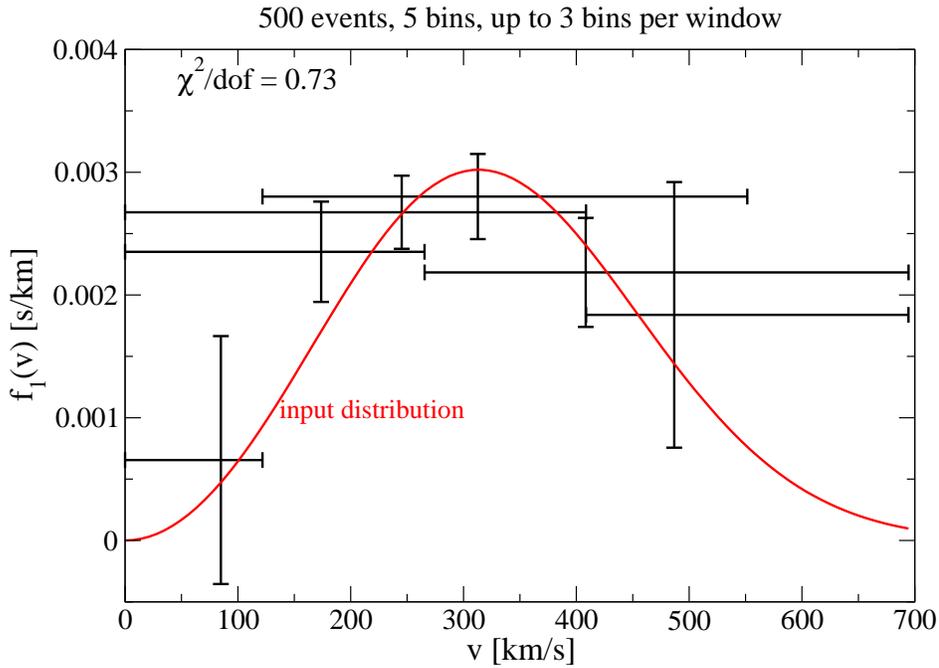}} \\
\vspace*{-5mm}
\rotatebox{270}{\includegraphics[width=10.5cm]{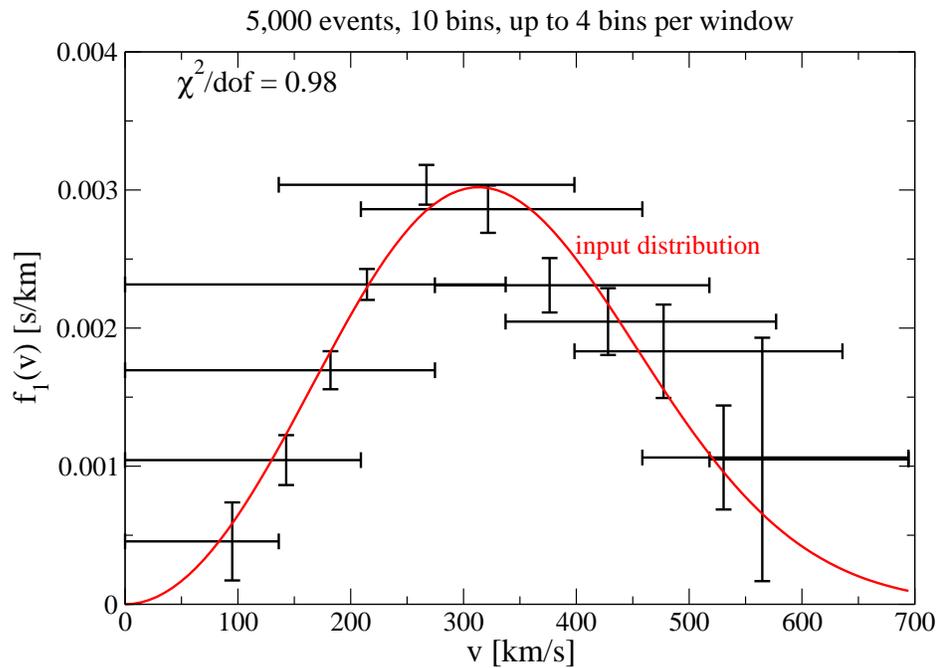}}
\end{center}
\vspace*{-5mm}
\caption{The WIMP velocity distribution reconstructed from a ``typical''
  experiment with 500 (top) and 5,000 (bottom) events. The smooth curves show
  the input distributions, which are based on Eq.(\ref{eqn222}). The vertical
  error bars show the square roots of the diagonal entries of the covariance
  matrix (\ref{e3_30}); the horizontal bars show the size of the window used
  in deriving the given value of $f_{1,r}$. The overlap of these horizontal
  bars thus shows the range over which the values of $f_{1,r}$ are
  correlated. Parameters as in Fig.~\ref{fig301}.}
\label{fig305}
\end{figure}

In Figs.~\ref{fig305} we show results for the reconstructed velocity
distribution, for ``typical'' simulated experiments with 500 (top) and 5,000
(bottom) events. In the top frame we choose $B=5$ bins, the first bin having a
width $b_1 = 8$ keV, and combine up to three bins into a window. Since the last
bin is in fact empty, this leaves us with $W=6$ windows, i.e., we can determine
$f_1$ for six discrete values of the WIMP velocity $v$; recall that these
``measurements'' of $f_1$ are correlated, as indicated by the horizontal bars
in the figure. In the lower frame we choose $B=10$ bins with $b_1 = 10$ keV,
and combine up to four bins into one window. The bins are thus significantly
smaller than in the upper frame. As a result, the last two bins are now
(almost) empty, leaving us with $W=11$ windows.

Figs.~\ref{fig305} indicate that one will need at least a few hundred events
for a meaningful direct reconstruction of $f_1$. Recall that $f_1$ is
normalized to unity. The overall magnitude of $f_1$ is therefore essentially
fixed by the range of observed $Q-$values; only the {\em shape} of this
distribution then remains to be determined. One measure of the information
content of the reconstructed $f_{1,r}$ is therefore the confidence level with
which one can exclude a constant $f_1$.

\begin{figure}[t]
\begin{center}
\rotatebox{270}{\includegraphics[width=13cm]{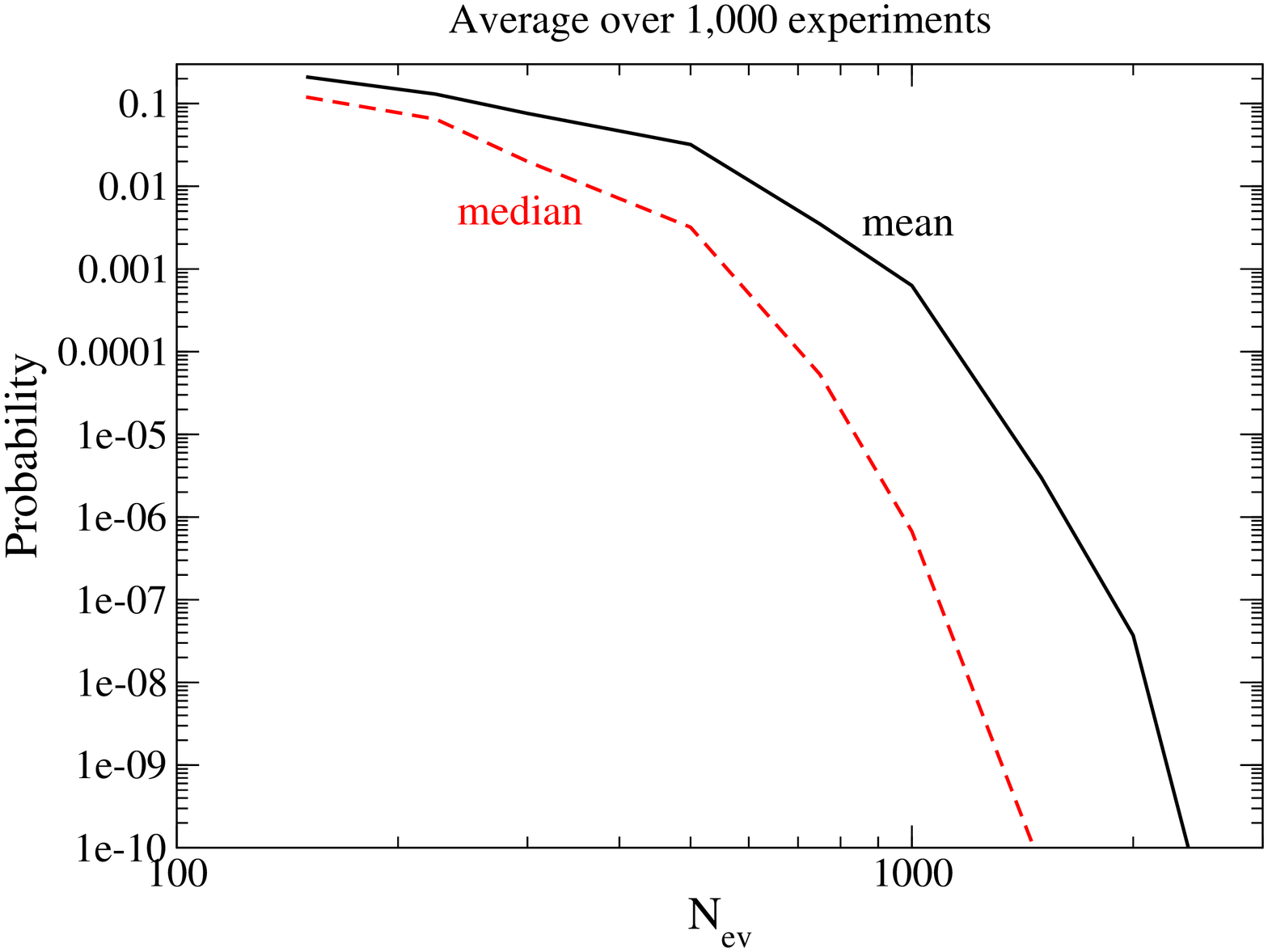}}
\end{center}
\caption{Estimates of the probability that the reconstructed velocity
  distribution is compatible with being constant, as a function of the average
  number of events per experiment. These results are for optimal combinations
  of $B, \ b_1$ and $n_W$; they have been obtained by averaging over 1,000
  simulated experiments. The solid (black) and dashed (red) curves show the
  mean and median values of the probability. Parameters as in Fig.~\ref{fig301}.}
\label{fig306}
\end{figure}

In Fig.~\ref{fig306} we show one minus this confidence level, i.e., the
probability that a reconstructed velocity distribution is compatible with a
constant. This has been estimated by defining a $\chi_f^2$ variable as in
Eq.(\ref{e3_31}) for the hypothesis $f_1 =$ const., and integrating the
theoretical $\chi^2$ distribution over the range $\chi^2 > \chi_f^2$. Here the
constant has been chosen as $1/(v_{\rm max} - v_{\rm min})$, where $v_{\rm
  min, max}$ have been calculated as in Eq.(\ref{eqn204}) using the largest
and smallest recoil energy, respectively, that has been measured in a given
experiment. Since this probability differs quite widely from one (simulated)
experiment to the next, we show both the mean and the median probability. We
see that we'll need at least 200 events if we want to reject the hypothesis
of a constant $f_1$ at the 90\% C. L. (on average). The confidence level then
increases very quickly as additional events are added; by the time 1,000
events have been accumulated, we can be quite sure that a constant $f_1$ can
be excluded with high confidence.

\begin{table}[t]
\begin{center}
\begin{tabular}{|c|c|c|c||c|c|c|c|}
\hline
$N_{rm ev}$ & $B$ & $b_1$ [keV] & $n_W$ & mean $P$ & median $P$ & $\langle
\Sigma \rangle $ & $\langle \chi_f^2 \rangle$ \\
\hline
500 & 5 & 30 & 3 & 0.032 & $3.2 \cdot 10^{-3}$ & 145 & 0.78 \\
500 & 5 & 30 & 2 & 0.08 & 0.02 & 107 & 0.77 \\
500 & 5 & 30 & 1 & 0.38 & 0.34 & 19 & 0.88 \\
500 & 5 & 30 & 4 & $9.0 \cdot 10^{-3}$ & $5.0 \cdot 10^{-4}$ & 165 & 1.3 \\
500 & 2 & 100 & 1 & 0.08 & 0.03 & 80 & 0.6 \\
\hline
5,000 & 6 & 20 & 3 & $2.8 \cdot 10^{-23}$ & $1.2 \cdot 10^{-40}$ & 1,360 &
1.03 \\
5,000 & 6 & 30 & 3 & $1.7 \cdot 10^{-26} $ & $2.0 \cdot 10^{-42}$ & 1,380 &
1.28 \\
5,000 & 10 & 10 & 4 & $2.9 \cdot 10^{-21}$ & $1.0 \cdot 10^{-37}$ & 1,520 &
0.85 \\
5,000 & 10 & 10 & 3 & $3.5 \cdot 10^{-11}$ & $2.5 \cdot 10^{-26}$ & 1,190 &
0.85 \\
5,000 & 10 & 10 & 2 & $9.5 \cdot 10^{-7}$ & $5.4 \cdot 10^{-12}$ & 490 &
0.88 \\
5,000 & 10 & 10 & 1 & 0.02 & $6.7 \cdot 10^{-4}$ & 48 & 0.93 \\
\hline
\end{tabular}
\end{center}
\caption{This table illustrates how the binning, and in particular the
  combination of bins into ``windows'', affects the information that can be
  gleaned from the reconstructed WIMP velocity distribution. The first four
  columns show the average number of events in a given experiment, the number
  of bins, the size of the first bin in keV, and the number of bins per
  window. The remaining four columns show the mean and median probability that
  the reconstructed $f_1$ is compatible with a constant, the mean of the
  quantity $\Sigma$ defined in Eq.(\ref{e3_32}), and the average $\chi_f^2$ of
  Eq.(\ref{e3_31}). }
\end{table}

This confidence level, as well as more general measures of the information
that can be extracted from a given experiment, depend on the choices of $B,
b_1$ and, in particular, $n_W$. This is illustrated in Table 1, which shows
results for different combinations of $B, \, b_1$ and $n_W$ for 500 (first
five rows) and 5,000 (last six rows) expected events per experiment. Here the
mean and median probabilities are the same as in
Fig.~\ref{fig306}.\footnote{The results of the first row in the Table have
  been entered in this figure.} In addition we show the mean of the quantity
$\Sigma$, defined as
\beq \label{e3_32}
\Sigma =  \sum_{\mu,\nu} {\cal C}_{\mu\nu} f_{1,r}(v_\mu) f_{1,r}(v_\nu) \, .  
\eeq
Formally $\Sigma$ determines the significance with which the hypothesis $f_1 =
0$ can be rejected. Since $f_1$ is normalized, this hypothesis is
unphysical. Nevertheless $\Sigma$ can be regarded as a measure of the
information content of a set of reconstructed $f_{1,r}(v_\mu)$; in the absence
of correlations, it becomes the sum over the inverse squares of the {\em
  relative} errors. Note that, in contrast to $\chi_f^2$, $\Sigma$ does not
have a $1/W$ factor in front; after all, by adding more windows we also add
more values $v_\mu$ at which $f_1$ is determined, which can increase the
information content.

The first four rows, as well as the last four rows, show the effect of varying
$n_W$, the maximal number of bins that are collected in a window. We see that
there is an optimal choice for this quantity. Reducing $n_W$ leads to loss of
information, as indicated by greatly increased values for the probability that
$f_{1,r}$ is compatible with $f_1$ being constant as well as reduced values of
$\Sigma$. On the other hand, making the windows too large introduces too large
systematic uncertainties in the estimates of the logarithmic slopes $k_\mu$,
which in turn leads to too large average values of $\chi_f^2$. This is
illustrated by rows four and seven, which have large windows due to our choice
of a large $n_W$ (row 4) or a large $b_1$ (row 7). 

The table also shows that the choice of $b_1$ has some impact on the
confidence level with which the hypothesis of a constant $f_1$ can be
rejected. We saw in Figs.~\ref{fig305} that our input $f_1$ has a broad maximum
at $v \simeq 300$ km/s. Rejection of the hypothesis of a constant $f_1$ is
therefore optimized by maximizing the information about the outer reaches of
$f_1$. Getting accurate information about $f_1$ at large velocities is very
difficult; this would need a large number of events at large $Q$, where the
counting rate is very small. This leaves the region of small WIMP velocity. By
choosing a large first bin, one greatly reduces the error on $f_{1,r}$ in this
first bin, which is also the first window; this was illustrated in
Fig.~\ref{fig302}. In fact, for 500 events and $n_W = 1$ one can formally
maximize the confidence level with which a constant $f_1$ can be rejected by
considering only two bins, and making the first bin very large; this is shown
in the fifth row. Note that this leads to an average $\chi_f^2$ well below
unity, indicating that in spite of the large bins, systematic errors are still
insignificant. However, we note that in this case our assumption that the
error on ${\cal N}$ is negligible is clearly not justified, since ${\cal N}$
receives almost its entire contribution from the large first bin. By including
the error on $r_1$ but ignoring the (strongly correlated!) error on ${\cal N}$
we clearly over--estimate the total statistical error in this case.
Recall also that a large first bin leads to a large value for the smallest
velocity, $v_1$,
where $f_1$ is determined.

Our ``figure of merit'' $\Sigma$ is less dependent on the details of binning,
although, as stated earlier, it does strongly benefit from combining several
bins into windows. We also note that the optimal achievable $\Sigma$ is
essentially proportional to the number of events in the sample. This is
expected, since $\Sigma$ is something like an inverse squared relative error. 

\subsection{Determining moments of \boldmath$f_1$}

We saw in the previous subsection that a direct reconstruction of the WIMP
velocity distribution $f_1$ will only be possible once several hundred elastic
nuclear recoil events have been collected. This is a tall order, given that
not a single such event has so far been detected (barring the possible DAMA
observation). The basic reason for the large required event sample is that,
$f_1$ being a normalized distribution, only information on the {\em shape} of
$f_1$ is meaningful. In order to obtain such shape information via direct
reconstruction, we have to separate the events into several bins or
windows. Moreover, each window should contain sufficiently many events to
allow an estimate of the {\em slope} of the recoil spectrum in this window.

On the other hand, at the end of Sec.~2 we also gave expressions for the {\em
  moments} of $f_1$. With the exception of the moment with $n=-1$, these are
entirely inclusive quantities, i.e., each moment is sensitive to the entire
data set; no binning is required, nor do we need to determine any slope (with
one possible minor exception; see below). It thus seems reasonable to expect
that one can obtain meaningful information about these moments with fewer
events.

An independent motivation for the determination of these moments is that they
are sensitive also to $f_1$ at large values of the WIMP velocity $v$. We saw
above that direct reconstruction of $f_1$ at large $v$ is very difficult, due
to the small number of events expected in this region. Moreover, a
delta--function--like contribution to $f_1$ at the highest velocity, $v = v_{\rm esc}$
is very difficult to detect using direct reconstruction; such a contribution
is expected in ``late infall'' models of galaxy formation \cite{modelc}.

The experimental implementation of Eq.(\ref{eqn215}) is quite straightforward.
For $Q_{\rm thre} = 0$, the normalization ${\cal N}$ has already been given in
Eq.(\ref{e3_29}). The case of non--vanishing threshold energy $Q_{\rm thre}$ can
be treated straightforwardly, using Eq.(\ref{en4}). To that end we need to
estimate the recoil spectrum at the threshold energy. One possibility would be
to choose an artificially high value of $Q_{\rm thre}$, and simply count the
events in a bin centered on $Q_{\rm thre}$. However, in this case the events
with $Q < Q_{\rm thre}$ would be left out of the determination of the moments.
We therefore prefer to keep $Q_{\rm thre}$ as small as experimentally possible,
and to estimate the counting rate at threshold using the ansatz
(\ref{eqn311}). Since we need the recoil spectrum only at this single point, we
only have to determine the quantities $r_1$ and $k_1$ parameterizing $dR/dQ$
in the first bin; this can be done as described in the previous subsection,
without the need to distinguish between bins and ``windows''. Introducing the
shorthand notation
\beq \label{e3_32a}
r_{\rm thre} \equiv \afrac {dR} {dQ}_{Q=Q_{\rm thre}}\, ,
\eeq
the resulting error can be written as
\beq \label{e3_33}
\sigma^2 (r_{\rm thre})  =
r_{\rm thre}^2 \cbrac{ \frac {\sigma^2(r_1)  }{r_1^2} + \left[ Q_{\rm thre} - Q_{s,1} -
    k_1 \afrac {\partial Q_{s,1}}  {\partial k_1} \right]^2 \sigma^2(k_1)} \, . 
\eeq
The squared errors for $r_1$ and $k_1$ are simply the corresponding diagonal
entries of the respective covariance matrices given in Eqs.(\ref{e3_23}) and
(\ref{e3_26}). Finally, the definition (\ref{e3_12}) of $Q_{s,1}$ implies
\beq \label{e3_34}
Q_{s,1} + k_1 \afrac {\partial Q_{s,1}} {\partial k_1} = Q_1 - \frac{1}{k_1} +
\afrac {b_1}{2} \coth x_1\, ,
\eeq
where $x_1 = b_1 k_1/2$ as before and $Q_1$ is the central $Q-$value in the
first bin. It should be noted that the first term in Eq.(\ref{eqn215}) is
negligible for all $n \geq 1$ if $Q_{\rm thre} \simeq 1$ keV; however, even for
this low threshold energy it contributes significantly to the normalization
constant ${\cal N}$, as described by Eq.(\ref{en4}). Of course, the first term
in Eq.(\ref{eqn215}) always dominates for $n = -1$. This is not surprising,
since the very starting point of our analysis, Eq.(1), already shows that the
counting rate at $Q_{\rm thre}$ is proportional to the ``minus first'' moment of
the velocity distribution.

The integral appearing in Eq.(\ref{eqn215}) can be estimated through the sum
\beq \label{e3_35}
I_n = \sum_a \frac {Q_a^{(n-1)/2}} {F^2(Q_a)} \, ,
\eeq
see Eq.(\ref{e3_29}). Since all $I_n$ are determined from the same data, they
are correlated, with
\beq \label{e3_36}
{\rm cov}(I_n, I_m) = \sum_a \frac {Q_a^{(n+m-2)/2}} {F^4(Q_a)} \, .
\eeq
This can e.g. be seen by writing Eq.(\ref{e3_35}) as a sum over narrow bins,
such that the recoil spectrum within each bin can be approximated by a
constant. Each term in the sum would then have to be multiplied with the
number of events in this bin; Eq.(\ref{e3_36}) then follows from standard
error propagation. Note that, when re--converted into an integral, the
expression for ${\rm cov}(I_0,I_0)$ will diverge logarithmically
 for $Q_{\rm thre} \rightarrow 0$; equivalently, the numerical estimate of this entry can
become very large if the sample contains events with very small
$Q-$values. Our numerical results presented below have therefore been obtained
with $Q_{\rm thre} = 1$ keV; many existing experiments in fact require
significantly larger energy transfers in their definition of a WIMP signal. 

We also need the correlation between the errors on the estimate of the
recoil spectrum at $Q = Q_{\rm thre}$ and the integrals $I_n$. It is clear that
these quantities are correlated, since the former is estimated from all events
in the first bin, which of course also contribute to the latter. We estimate
these correlations again using the ansatz (\ref{eqn311}), which makes the
following prediction for the contribution of the first bin to the integrals:
\beq \label{e3_38}
I_{n,1} = r_1 \int_{Q_{\rm thre}}^{Q_{\rm thre}+b_1} \bfrac {Q^{(n-1)/2}} {F^2(Q)}
{\rm e}^{k_1 (Q - Q_{s,1})} \, dQ \, .
\eeq
This immediately implies $\partial I_{n,1} / \partial r_1 = I_{n,1} / r_1$,
and
\beq \label{e3_39}
\frac {\partial I_{n,1}} {\partial k_1} = I_{n+2,1} - \left[ Q_{s,1} + k_1
  \afrac {\partial Q_{s,1}} {\partial k_1} \right] I_{n,1}\, 
\eeq
see Eq.(\ref{e3_34}). Note that the $I_{n,1}$ and $I_{n+2,1}$ in Eq.(\ref{e3_39}) are
evaluated as in Eq.(\ref{e3_35}), with the sum extending only over events in
the first bin. The correlation we're after is then given by
\beqn \label{e3_40}
\conti {\rm cov} ( r_{\rm thre}, I_n ) \non\\
 \= 
r_{\rm thre} \~ I_{n,1} \left\{ \frac  {\sigma^2(r_1)} {r_1^2}
+ \left[ \frac{I_{n+2,1}}{I_{n,1}} - Q_{s,1} - k_1 \afrac {\partial Q_{s,1}} {\partial k_1} \right]
  \left[ Q_{\rm thre} - Q_{s,1} - k_1 \afrac {\partial Q_{s,1}} {\partial k_1} \right]\!
 \sigma^2(k_1) \right\} \, .
\nonumber\\
\eeqn
These ingredients allow us to compute the covariance matrix for our estimates
of the moments of the velocity distribution: 
\begin{eqnarray} \label{e3_41}
\conti {\rm cov} \aBig{\langle v^n \rangle, \langle v^m \rangle} \non\\
\= 
{\cal N}_{\rm m}^2 \bigg[ \langle v^n \rangle \langle v^m \rangle {\rm cov}
  (I_0,I_0) + \alpha^{n+m} (n+1) (m+1) {\rm cov}(I_n, I_m) 
\nonumber \\ 
\conti ~~~~~~~~ 
 -\alpha^m (m+1) \langle v^n \rangle {\rm cov}(I_0, I_m)
 -\alpha^n (n+1) \langle v^m \rangle {\rm cov}(I_0, I_n)
\nonumber \\
\conti ~~~~~~~~~~~~~~ 
 + D_n D_m \sigma^2 ( r_{\rm thre} ) 
 - \Big( D_m \langle v^n \rangle + D_n  \langle  v^m \rangle
 \Big) {\rm cov} \left (r_{\rm thre},I_0 \right) \Bigg.
\nonumber \\ 
\conti ~~~~~~~~~~~~~~~~~~~~ 
+ \alpha^m (m+1) D_n {\rm cov}(r_{\rm thre}, I_m)
 + \alpha^n (n+1) D_m {\rm cov} (r_{\rm thre}, I_n)  \bigg] \, .
\end{eqnarray}
Here we have introduced the modified normalization
\beq \label{e3_42}
{\cal N}_{\rm m} \equiv \afrac{\alpha}{2} {\cal N}\, ,
\eeq
which exploits the partial cancellation of the $\alpha$ factors between
Eqs.(\ref{eqn215}) and (\ref{en4}), and the quantities
\beq \label{e3_43}
D_n \equiv \frac {1} {{\cal N}_{\rm m}} 
\afrac {\partial \langle v^n \rangle} {\partial r_{\rm thre}} 
= \frac{2} {F^2(Q_{\rm thre})} \left( \alpha^n Q_{\rm thre}^{(n+1)/2} -
  \sqrt{Q_{\rm thre}} \~ \langle v^n \rangle \right) \, .
\eeq

In our numerical simulations we find that Eq.(\ref{e3_35}) indeed reproduces
the ``exact'' (input) values of the $I_n$ if one includes sufficiently many
events. In this case it does not matter whether one considers a single
experiment with a large number of events, or averages over many simulated
experiments with a relatively small number of events. However, in the second
case the average values of the reconstructed moments do not exactly converge
to the input values. In order to understand this, consider the simple case
$Q_{\rm thre} = 0$. The moments are then proportional to the ratio $I_n / I_0$,
see Eqs.(\ref{eqn215}) and (\ref{en4}). The distortion arises because $\langle
I_n / I_0 \rangle \neq \langle I_n \rangle / \langle I_0 \rangle$, where the
averaging is over many simulated experiments. In Appendix B we show how this
can be corrected using Taylor expansion to second order. The leading correction terms
for small $Q_{\rm thre}$ and not very large first bin are
\begin{eqnarray} \label{e3_44}
\delta \langle v^n \rangle
 \= \alpha^n {\cal N}_{\rm m}^2  \Bigg\{  (n+1) \bigg[ {\rm
  cov}(I_0, I_n) - {\cal N}_{\rm m} I_n {\rm cov}(I_0, I_0) \bigg]
\nonumber \\
\conti ~~~~~~~~~~~~ 
+ 2 \bfrac {Q_{\rm thre}^{(n+1)/2}} {F^2(Q_{\rm thre})} \bigg[ {\rm cov}( r_{\rm thre}, I_0)
 - r_{\rm thre} {\cal N}_{\rm m} {\rm cov}(I_0,I_0) \bigg] \Bigg\} \, ,
\end{eqnarray}
 the second line is significant only for $n = -1$.
Note that this correction becomes very small if the statistical errors on the
$I_n$ as well as on $r_{\rm thre}$ become small.

With this correction, the reconstructed $\langle v^n \rangle$ indeed closely
reproduce the input values after averaging over sufficiently many experiments,
even if the number of events in a given experiment is small. However, the
numerical analysis revealed a number of additional problems. These can be
understood from the observation that the $I_n$ in Eq.(\ref{e3_35}), and even
more the entries in their covariance matrix (\ref{e3_36}), receive significant
contributions from large $Q$ values, where the counting rate itself is already
very small. 

\begin{figure}[t]
\begin{center}
\rotatebox{270}{\includegraphics[width=13cm]{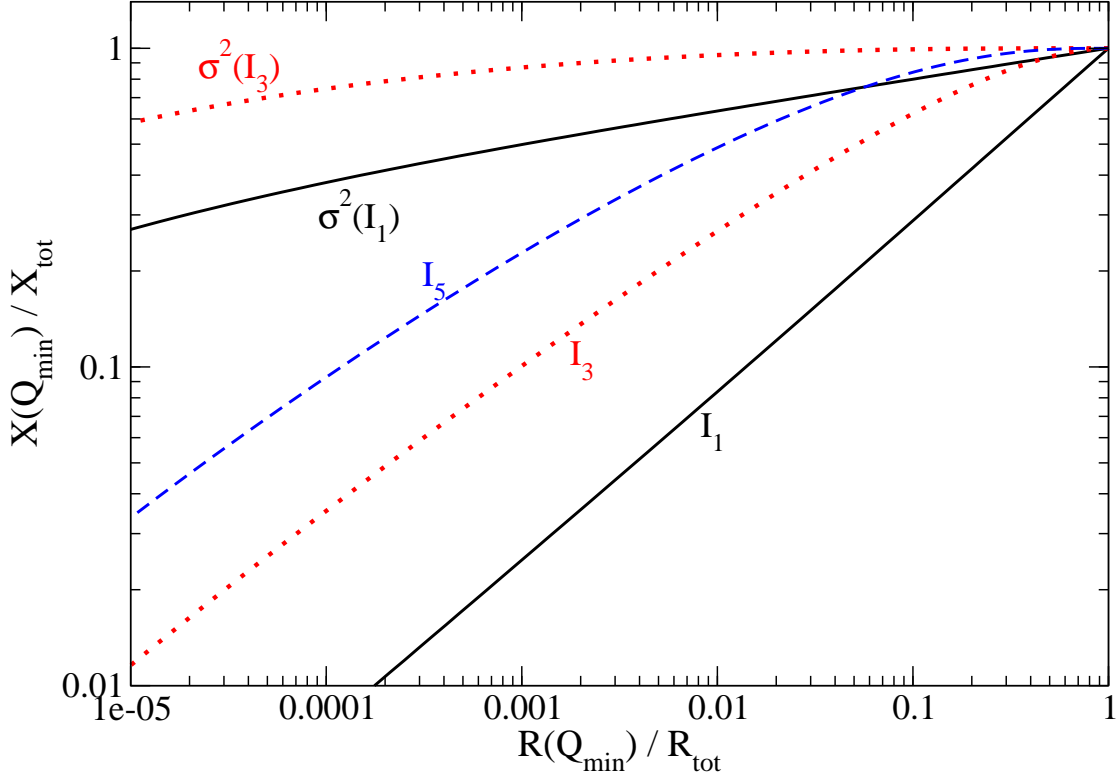}}
\end{center}
\caption{This figure has been obtained by introducing an artificial lower
  bound $Q_{\rm min}$ in the counting rate as well as in the definition of the
  integrals $I_n$. The $x-$axis shows the counting rate for different $Q_{\rm
  min}$, where small $R(Q_{\rm min})$ corresponds to large $Q_{\rm min}$. The
  $y-$axis shows the impact of varying $Q_{\rm min}$ on $I_1$ (solid, black),
  $I_3$ (dotted, red) and $I_5$ (dashed, blue); the upper curve of a given
  pattern refers to ${\rm cov} (I_n,I_n)$. The values of the parameters are as
  in Fig.~\ref{fig301}. See the text for further details.}
\label{fig307}
\end{figure}

This is illustrated in Fig.~\ref{fig307}. The $x-$axis shows the quantity
\beq \label{e3_45}
R(Q_{\rm min}) \equiv \int_{Q_{\rm min}}^{Q_{\rm max}} \afrac {dR} {dQ} dQ\, ,
\eeq
divided by the total counting rate $R \equiv R(Q_{\rm min} = 0)$. Here,
$Q_{\rm max}$ is the kinematic maximum of $Q$ for given input parameters, and
$Q_{\rm min}$ is varied freely between 0 and $Q_{\rm max}$. The $y-$axis shows
analogously the contributions to some $I_n$ (lower curves) and to the
corresponding diagonal elements of the covariance matrix, i.e., the squared
errors (upper curves), that come from the region $Q > Q_{\rm min}$. In the
latter case we have converted the sums in Eq.(\ref{e3_36}) back into
integrals. The figure shows that the region of $Q-$values that contributes
99\% of the counting rate only contributes about 92\% to $I_1$, 73\% to $I_3$
and 51\% to $I_5$; for the given input parameters, this corresponds to the
region $Q \leq 103$ keV. Even worse, this region only contributes about 35\%
to ${\rm cov}(I_1,I_1)$ and 5\% to ${\rm cov}(I_3,I_3)$! This implies that
an experiment collecting only a small number of events will typically
underestimate $\langle v^n \rangle$ and, especially, its error; the problem
will become worse with increasing $n$. On the other hand, as mentioned above,
when averaged over sufficiently many experiments, our estimates for the $I_n$ 
do reproduce the true (input) values. This implies that occasionally an
experiment will greatly {\em over}estimate the $I_n$, the problem again
getting worse for larger $n$.

Our numerical analysis also shows that, after averaging over (very) many
experiments, Eq.(\ref{e3_36}) reproduces the mean square deviation between our
estimated $I_n$ and the true (input) value. Nevertheless, we just saw that in
most cases this error is being underestimated. In order to be conservative, we
therefore added ``the error on the error'' to the diagonal entries of the
covariance matrix; the off--diagonal entries are then scaled up such that the
correlation matrix remains unaltered. The squared ``error on the error'' is
defined as
\beq \label{e3_46}
\sigma^2\big({\rm cov}(I_n,I_n) \big) = \sum_a \frac {Q_a^{2n-2}} {F^8(Q_a)}
\, .
\eeq 
With this modification, the average $\chi^2$, again averaged over many
experiments, is in the vicinity of unity at least for the first few
moments.\footnote{Note that $\langle {\rm cov}(I_n,I_n) \rangle = \langle (
  I_n - I_{n,r})^2 \rangle$, where $I_{n,r}$ are the values of our estimators
  based on Eq.(\ref{e3_35}) and $I_n$ are the true (input) values, in general
  does not imply that $\langle {\rm cov}(I_n,I_n) / ( I_n - I_{n,r})^2 \rangle
  = 1$. Adding the ``error of the error'' to the covariance matrix brings the
  average of this ratio closer to unity.}

Another problem is that the errors of the $I_n$ are very highly correlated.
This can also be understood from Fig.~\ref{fig307}: A single event at high $Q$
will contribute greatly to all moments with sufficiently large $n$.
Numerically we find correlations of more than 98\% between $\langle v^n
\rangle$ and $\langle v^{n+1} \rangle$ for all $n \geq 2$; the correlation
between $\langle v \rangle$ and $\langle v^3 \rangle$ still amounts to more
than 87\%.  This implies that the higher moments unfortunately add only little
to the available information. Worse, attempting to include high moments in a
$\chi^2$ fit often leads to numerical instabilities; recall that a covariance
matrix containing 100\% correlated entries become singular, i.e., can no longer
be inverted. In practice only the moments with $n \~ \lsim \~ 3$ therefore seem to
be useful.

\begin{figure}[t]
\begin{center}
\rotatebox{270}{\includegraphics[width=10cm]{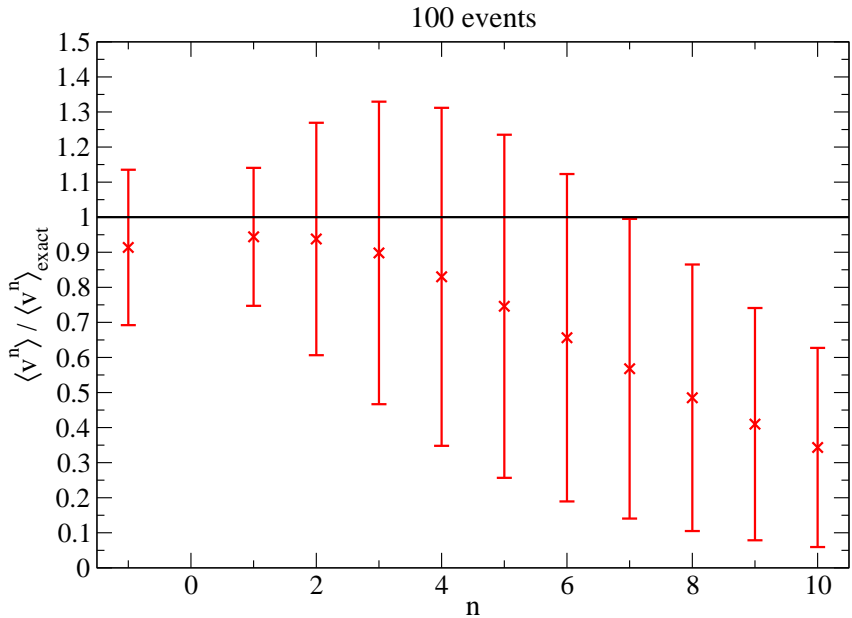}}
\end{center}
\caption{Estimated moments, and their errors, for a ``typical'' simulated
  experiment with 100 events; recall that the errors are strongly
  correlated. Parameters are as in Fig.~\ref{fig301}.}
\label{fig308}
\end{figure}

We are now ready to present some representative numerical results.
Fig.~\ref{fig308} shows the first 10 moments reconstructed with 100 events,
using our standard input parameters (see Fig.~\ref{fig301}). The estimated values of the
moments have been divided by the true values. We see that in this ``typical''
example the high moments are indeed underestimated.  We also see that the
estimated relative errors at first increase with increasing $n$; this
reproduces correctly the trend of the actual deviation of the estimated
moments from the exact values. However, even after adding ``the error on the
error'', we find that the relative errors start to decline again for $n>6$.
This effect is probably entirely spurious; recall that the errors are likely
to be even more underestimated than the moments themselves. Nevertheless we
find it encouraging that already with 100 events a couple of moments can be
determined with errors of about 15\%.

\begin{figure}[t!]
\begin{center}
\rotatebox{270}{\includegraphics[width=10.5cm]{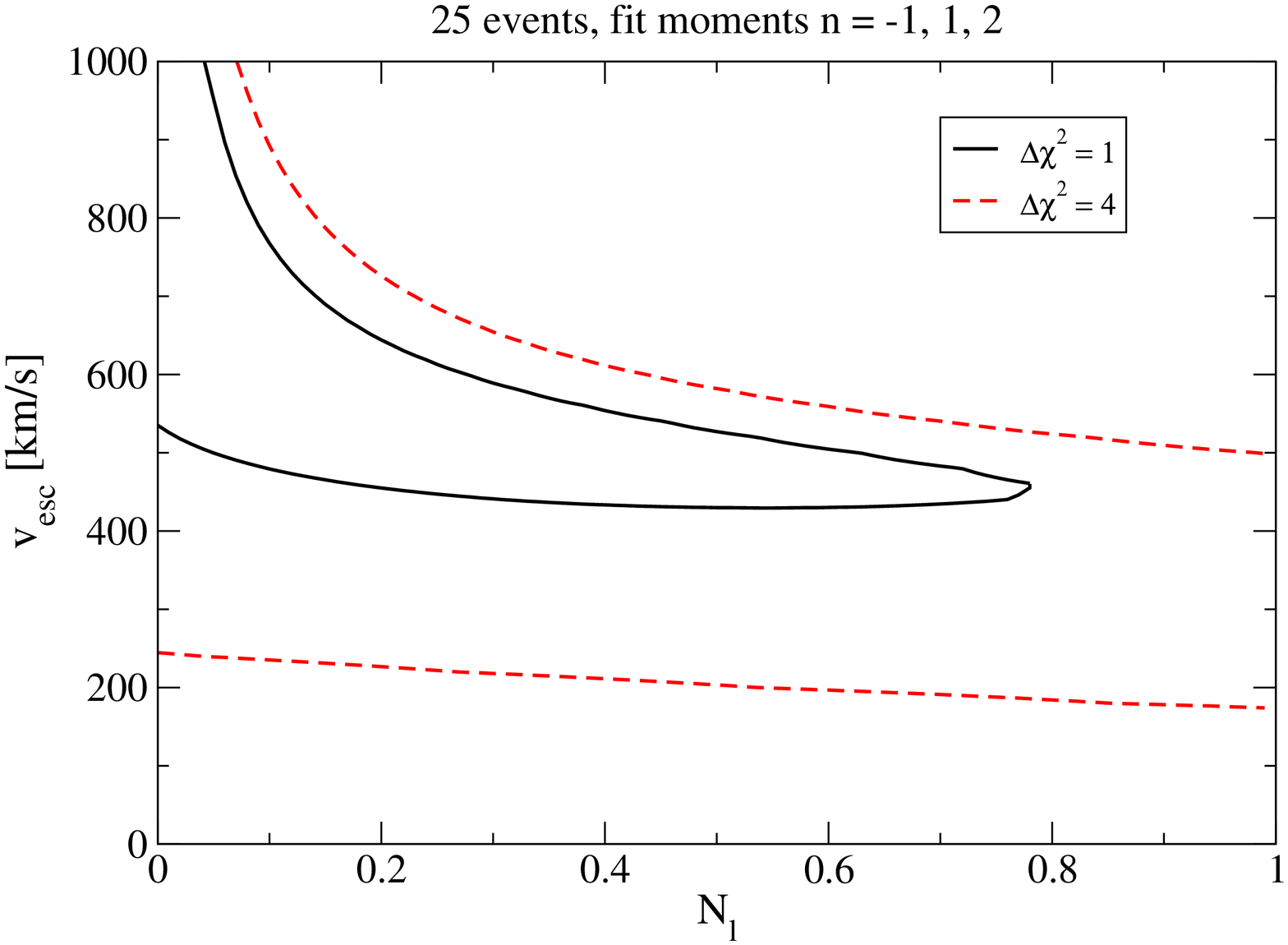}}
\vspace*{-5mm}
\rotatebox{270}{\includegraphics[width=10.5cm]{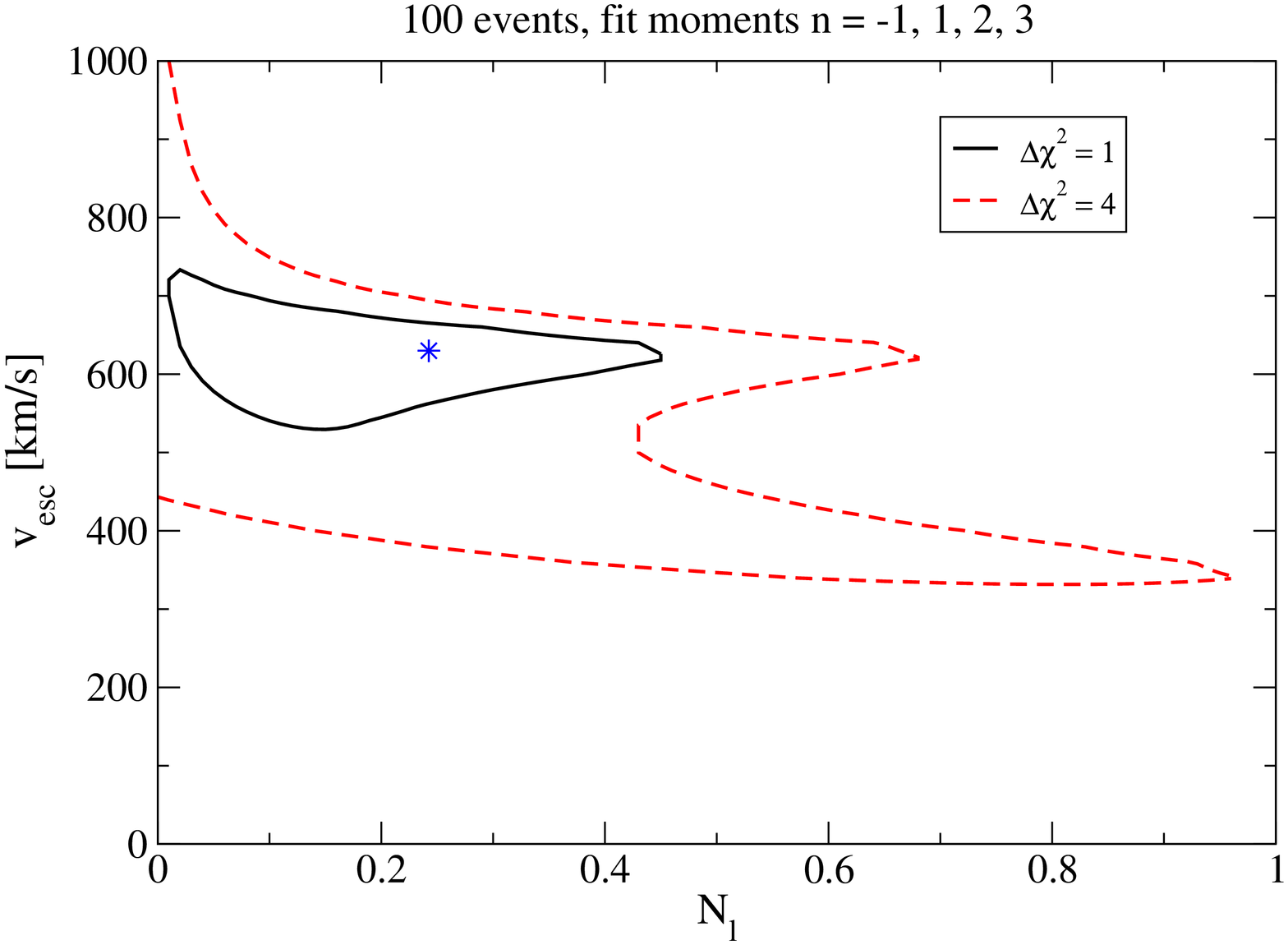}}
\end{center}
\vspace*{-5mm}
\caption{$\chi^2$ contours, calculated from 3 (top) and 4 (bottom) moments, in
  the plane spanned by the normalization of the ``late infall'' component and
  the galactic escape velocity, for typical experiments with 25 (top) and 100
  (bottom) events. In the upper frame the minimal $\chi^2$ value is close to
  the input values $N_l = 0, \, v_{\rm esc} = 700$ km$/$s; in the lower frame, the
  location of the minimal $\chi^2$ is indicated by the star. Parameters are as
  in Fig.~\ref{fig301}.}
\label{fig309}
\end{figure}

Figs.~\ref{fig309} show a example of the information that might be gleaned
from analyses of reconstructed moments of the WIMP velocity distribution. Here
we attempt to constrain a possible ``late infall'' component in $f_1$
\cite{modelc}, defined by the ansatz
\beq \label{e3_47}
f_1(v; v_{\rm esc}, N_l) = N_s f_{1,\sh}(v) \theta(v_{\rm esc} - v) + N_l \delta(v
- v_{\rm esc}) \, .
\eeq
Here $f_{1,\sh}$ is the standard ``shifted Gaussian'' distribution (\ref{eqn222}).
As before, we have multiplied it with a
cut--off at $v_{\rm esc}$. In addition, we introduce a contribution of WIMPs with
fixed velocity, which we set equal to $v_{\rm esc}$; these WIMPs are just falling into
our galaxy.\footnote{Strictly speaking this distribution should also be
  smeared, since the velocity of the Earth relative to the galaxy can add or
  subtract to the infall velocity. However, as long as $v_{\rm esc}$ is significantly
  larger than $v_e$, this smearing should not matter very much.
 See Ref.~\cite{Ling04} for a discussion of the effect of late WIMP
 infall on the recoil spectrum.} $N_l$ and
$v_{\rm esc}$ are our free parameters; $N_s$ is then chosen such that the total $f_1$
is normalized to unity. We then plot contours of $\delta \chi^2$, defined as
the deviation of $\chi^2$ from its minimal value, where $\chi^2$ is defined as
\beq \label{e3_48}
\chi^2 = \sum_{m,n} \left[ \langle v^m \rangle_r - \langle v^m\rangle(v_{\rm esc},N_l)
\right] {\cal D}_{mn} \left[ \langle v^n \rangle_r - \langle v^n\rangle(v_{\rm esc},
  N_l) \right] \, ;
\eeq
here $\langle v^n \rangle_r$ are the reconstructed moments in our mock
experiment, $\langle v^n \rangle(v_{\rm esc},N_l)$ are the predictions for these
quantities based on Eq.(\ref{e3_47}), and ${\cal D}$ is the inverse of the
covariance matrix (\ref{e3_41}).

Figs.~\ref{fig309} show a strong degeneracy in the fit. If the galactic escape
velocity $v_{\rm esc}$ is kept fixed at its input value of 700 km/s, a quite
significant upper bound on the normalization $N_l$ of the late infall
component could be derived already from our simulated experiment with 25
events. However, if $v_{\rm esc}$ is kept free, no significant upper bound can be
derived even from the simulated experiment with 100 events. Note that the two
experiments have been simulated independently, i.e., the 25 events used for the
analysis in the upper frame are not part of the 100 events used in the lower
frame. The simulated experiment with 100 events was a bit ``unlucky'' in that
the input values lie just outside the $\Delta \chi^2 = 1$ contour. As a
result, the upper bound on $N_l$ for fixed $v_{\rm esc} = 700$ km/s actually comes out
a little worse in this case than in the experiment with only 25 events. This
is in spite of the fact that the larger data sample allowed us to include one
more moment in the fit.

Note that, according to the definition (\ref{e3_47}), all WIMPs in our
galactic neighborhood have velocity $v < v_{\rm esc}$. This implies that a lower
bound on $v_{\rm esc}$ can be derived from the highest observed $Q-$value, $v_{\rm esc} >
\alpha \sqrt{Q_+}$, see Eq.(\ref{eqn204}). Unfortunately for our standard set
of input parameters, this method on average only yields lower bounds of about
400 (460) km/s for experiments with 25 (100) events. Even the experiment with
100 events would then still allow 60\% of all WIMPs to originate from a late
infall component; this is to be compared with the theoretical expectation
$N_l \~ \lsim \~ 10\%$. Finally, we note that for $N_l = 0$, it will be essentially
impossible to derive a meaningful upper bound on $v_{\rm esc}$ from these experiments:
because the original ``shifted Gaussian'' velocity distribution is already
very small at our input value $v_{\rm esc} = 700$ km/s, increasing $v_{\rm esc}$ has
essentially no effect on the measured recoil spectrum.

\section{Summary and Conclusions}

In this paper, we have developed methods that allow to extract information on
the WIMP velocity distribution from the recoil energy spectrum $dR/dQ$
measured in elastic WIMP--nucleus scattering experiments. In the long term
this information can be used to test or constrain models of the dark halo of
our galaxy; this information would complement the information on the density
distribution of WIMPs, which can be derived e.g. from measurements of the
galactic rotation curve.

To this end, in Sec.~2 we derived expressions that allow to directly
reconstruct the normalized one--dimensional velocity distribution function of
WIMPs, $f_1(v)$, given an expression (e.g., a fit to data) for the recoil
spectrum. We have also derived formulae for the moments of $f_1$. All these
expressions are independent of the as yet unknown WIMP density near the Earth
as well as of the WIMP--nucleus cross section; the only information about the
nature of the WIMP that is needed is its mass.

Furthermore, in Sec.~3 we have developed methods that allow to apply our
expressions directly to data, without the need to fit the recoil spectrum to a
functional form. We found that a good variable that allows direct
reconstruction of $f_1$ is the average recoil energy in a given bin (or
``window''; see Sec.~3.2). This average energy is sensitive to the slope of
the recoil spectrum, which is the quantity we need to reconstruct $f_1$. We
carefully analyzed the statistical errors. Unfortunately we found that several
hundred events will be needed for this method to be able to extract meaningful
information on $f_1$. This is partly due to the fact that $f_1(v)$ is
normalized, i.e., only the {\em shape} of this distribution contains meaningful
information, and partly because this shape depends on the {\em slope} of the
recoil spectrum, which is intrinsically difficult to determine.

We therefore turned to an analysis of the moments of $f_1$. We found that
reliably estimating higher moments, and in particular estimating their errors,
is difficult. The main reason is that these higher moments get large
contributions from very rare events with large recoil energies. Nevertheless
we found that, based only on the first two or three moments, some non--trivial
information can already be extracted from ${\cal O}(20)$ events.

The main emphasis of this exploratory study was on the basic expressions as
well as on their implementation in actual experiments. The models for $f_1$ we
tried to constrain in our applications (a constant in Sec.~3.2, a
``late--infall'' component with fixed velocity in the Earth rest frame in
Sec.~3.3) are not physical; nevertheless they illustrate the difficulties one
will have in extracting information from these experiments that are of
interest for the modeling of the galactic WIMP distribution.

Our analysis is based on several simplifying assumptions. First, we ignored
all experimental systematic uncertainties, as well as the uncertainty on the
determination of the recoil energy $Q$. This is probably quite a good
approximation, given that we found that we'll have to live with quite large
statistical uncertainties in the foreseeable future; recall that not a single
WIMP event has as yet been unambiguously recorded.

We also assumed that our detector consists of a single isotope. This is quite
realistic for the current semiconductor (Si or Ge) detectors. The CRESST
detector \cite{CRESST} contains three different nuclei. However, by
simultaneously measuring heat and light, one might be able to tell on an
event--by--event basis which kind of nucleus has been struck. In this case,
our methods can be applied straightforwardly to the three separate sub--spectra.

Our analyses treat each recorded event as signal, i.e., we ignore backgrounds
altogether. At least after introducing a lower cut $Q_{\rm thre}$ on the recoil
energy, this may in fact not be unrealistic for modern detectors, which
contain cosmic ray veto and neutron shielding systems. Background subtraction
should be relatively straightforward when fitting some function to the data,
which would allow to use the expressions of Sec.~2. It should also be feasible
in the method described in Sec.~3.2, if its effect on the average $Q-$values
in the bins can be determined; in particular, an approximately flat
($Q-$independent) background would not change the slope of the recoil
spectrum. Subtracting the background in the determination of the moments as
described in Sec.~3.3 might be (even) more difficult.

As noted earlier, we need to know the WIMP mass $m_\chi$. This is true for any
reconstruction method based on data taken with a mono--isotopic detector. In
this case one can always ``reconstruct'' $f_1(v)$, for any (assumed) value of
$m_\chi$. Fortunately in well--motivated WIMP models, $m_\chi$ can be
determined with high accuracy from future collider data. Even in this case one
will want to check experimentally that the WIMPs seen in Dark Matter detection
experiments are in fact the same ones produced at colliders. This can be done
by using the methods developed here on two different data sets, obtained with
different detector materials, and demanding consistent results for (the
moments of) $f_1$. The feasibility of such an analysis is currently under
investigation.

In our analysis we ignored the annual modulation of the WIMP flux. Again,
given the large statistical errors expected in the foreseeable future, this is
a reasonable first approximation. Nevertheless, eventually one will have to 
extend the formulae and methods developed here to allow for an annual
modulation. This is fairly straightforward if the background is again
negligible. On the other hand, new methods may be needed to extract
information on $f_1$ in a situation where the total counting rate is dominated
by background events; this is most likely the case for the DAMA data
\cite{DAMA}, even if they indeed contain a signal, which remains highly
controversial. 

In summary, we have begun to explore what direct Dark Matter detection
experiments can teach us about the velocity distribution of Dark Matter
particles in our galactic neighborhood. Our analyses show that this will
require substantial data samples. We hope this will encourage our experimental
colleagues to plan future experiments well beyond the stage of ``merely''
detecting Dark Matter.

\subsubsection*{Acknowledgments}
We thank Holger Drees for illuminating discussions on stochastics. This work
was partially supported by the Marie Curie Training Research Network
``UniverseNet'' under contract no.  MRTN-CT-2006-035863, as well as by the
European Network of Theoretical Astroparticle Physics ENTApP ILIAS/N6 under
contract no.  RII3-CT-2004-506222.

\appendix
\setcounter{equation}{0}
\setcounter{figure}{0}
\renewcommand{\theequation}{A\arabic{equation}}
%
%
\section{Normalization constant and moments of \boldmath$f_1$}
\label{appN}

Since
\beq
v = \alpha \sqrt{Q},
\label{eqnA01}
\eeq
we have
\beq
dv = \afrac{\alpha}{2 \sqrt{Q}} dQ,
\label{eqnA02}
\eeq
From Eq.(\ref{eqn212}) and according to the normalization condition in
Eq.(\ref{eqn214}), we have,
\beqn
\intz f_1(v) \~ dv \&=\& \calN \intz \left\{-2 Q \cdot \ddRdQoFQdQ\right\}
\afrac{\alpha}{2 \sqrt{Q}} dQ 
       \nonumber\\
 \&=\& \calN \cdot \abrac{-\alpha} \intz \sqrt{Q} \cdot \ddRdQoFQdQ dQ
       \nonumber\\
 \&=\& \calN \cdot \abrac{-\alpha} \cbrac{\sqrt{Q} \bdRdQoFQ_0^{\infty} -
   \frac{1}{2} \intz \frac{1}{\sqrt{Q}} \bdRdQoFQ dQ}   \nonumber\\
 \&=\& \calN \afrac{\alpha}{2} \intz \frac{1}{\sqrt{Q}} \bdRdQoFQ dQ
       \nonumber\\
 \&=\& 1,
\label{eqnA03}
\eeqn
where we have used the conditions
\beq
\dRdQ\Bigg|_{Q \to \infty} \to 0
\eeq
and
\beq
\dRdQ\Bigg|_{Q \to 0} \neq \infty.
\eeq
Eq.(\ref{eqn213}) immediately followed from Eq.(\ref{eqnA03}). 

Using Eqs.(\ref{eqnA01}), (\ref{eqnA02}) and integration by parts, we can also
find the moments of $f_1$, defined with a lower cut--off $Q_{\rm thre}$ on the
energy transfer, as follows:
\beqn
\expv{v^n} \&=\& \int_{v_{\rm min}(Q_{\rm thre})}^\infty v^n f_1(v) \~ dv
\nonumber\\ 
\&=\& \calN \int_{Q_{\rm thre}}^\infty \aBig{\alpha \sqrt{Q}}^n \cbrac{-2 Q
  \cdot \ddRdQoFQdQ} \afrac{\alpha}{2 \sqrt{Q}} dQ        \nonumber\\
 \&=\& \calN \cdot \abrac{-\alpha^{n+1}} \int_{Q_{\rm thre}}^\infty
 Q^{(n+1)/2} \cdot \ddRdQoFQdQ dQ       \nonumber\\
 \&=\& \calN \alpha^{n+1} \left\{ \frac {Q_{\rm thre}^{(n+1)/2}}
     {F^2(Q_{\rm thre})} \adRdQ_{Q=Q_{\rm thre}} + \frac{n+1}{2} \int_{Q_{\rm thre}}^\infty  Q^{(n-1)/2}
     \bdRdQoFQ dQ \right\} \, .
\eeqn
This reproduces Eq.(\ref{eqn215}) in Sec.~2.


\section{Derivation of the correction terms in Eq.(\ref{e3_44})}
\label{corr}
Starting point is the observation that we wish to compute the ratio of two
integrals,
\beq \label{eb1}
\frac {G_1}{G_2} = \frac {\int g_1(x) \~ dx} {\int g_2(x) \~ dx}
\rightarrow \frac {\sum_i n_i g_1(x_i) } {\sum_j n_j g_2(x_j)} \, .
\eeq
In the second step the integrals have been discretized, i.e., replaced by sums
over bins $i$ with $n_i$ events per bin. We now write the $n_i$ as sum of
average value $\bar n_i$ and fluctuation $\delta n_i$:
\beq \label{eb2}
\frac {G_1}{G_2} = \frac { \sum_i (\bar n_i + \delta n_i) g_1(x_i) } {\sum_j \bar n_j
  g_2(x_j) + \sum_j \delta n_j g_2(x_j) } \, .
\eeq
Introducing the notation
\beq
 \Bar G_a = \sum_i \bar n_i g_a(x_i)\,, ~~~~~~~~~~~~~~~~~~~~~~~~ a = 1,~2, 
\eeq
and expanding up to second order in the $\delta n_i$, we have:
\begin{eqnarray} \label{eb3}
\frac {G_1}{G_2}  \&\simeq\& \frac {\Bar G_1 + \sum_i \delta n_i g_1(x_i) } {\Bar G_2} \left[
1 - \frac {\sum_j \delta n_j g_2(x_j) } {\Bar G_2} + \left(
\frac {\sum_j \delta n_j g_2(x_j) } {\Bar G_2} \right)^2 \right]
\nonumber \\
\&\simeq\& \frac {\Bar G_1} {\Bar G_2} + \frac {1}{\Bar G_2} \abrac{\sum_i \delta n_i g_1(x_i)}
 - \frac {\Bar G_1} {\Bar G_2^2} \abrac{\sum_i \delta n_i g_2(x_i)}
\nonumber \\ 
\&~\& ~~~~~~~~ 
- \frac{1}{\Bar G_2^2} \abrac{\sum_i \delta n_i g_1(x_i)} \aBigg{\sum_j \delta n_j g_2(x_j)}
 + \frac {\Bar G_1} {\Bar G_2^3} \left( \sum_i \delta n_i g_2(x_i) \right)^2
\, .
\end{eqnarray}

We now consider the average over many experiments. Of course, $\delta n_i$
averages to zero, but the product $\delta n_i \delta n_j$ averages to $\bar
n_i \delta_{ij}$, i.e., it is non--zero for $i = j$. Hence:
\beq \label{eb4}
 \Expv{\frac {G_1}{G_2}} \simeq \frac {\Bar G_1} {\Bar G_2} 
- \frac{1}{\Bar G_2^2} \abrac{ \sum_i \bar n_i g_1(x_i) g_2(x_i) }
+ \frac {\Bar G_1} {\Bar G_2^3} \abrac{\sum_i \bar n_i g_2^2(x_i)} \, .
\eeq
The sums appearing in the two correction terms also appear in the definition
of the covariance matrix between $G_1$ and $G_2$. Note that we wish to compute
the first term on the right--hand side, since in our case the estimators for
$G_1$ and $G_2$ indeed average to the correct values. This then leads to the
final result
\beq \label{eb5}
 \frac {\Bar G_1} {\Bar G_2} -  \Expv{\frac {G_1}{G_2}} 
= \afrac{1}{\Bar G_2^2} {\rm cov}(G_1,G_2)
 - \afrac{ \Bar G_1} {\Bar G_2^3} {\rm cov}(G_2,G_2) \, .
\eeq
Applying this result to Eqs.(\ref{eqn215}) and (\ref{en4}) then immediately
leads to Eq.(\ref{e3_44}).


\begin{thebibliography}{99}
%
\bibitem{evida}
F. Zwicky, {\it Helv. Phys. Acta} {\bf 6}, 110 (1933); 
S. Smith, {\it Astrophys. J.} {\bf 83}, 23 (1936).

\bibitem{evidb}
V. C. Rubin and W. K. Ford, {\it Astrophys. J.} {\bf 159}, 379 (1970);
S. M. Faber and J. S. Gallagher, {\it Annu. Rev. Astron. Astrophys.} {\bf 17}, 135 (1979);
V. C. Rubin, W. K. Ford, and N. Thonnard, {\it Astrophys. J.} {\bf 238}, 471 (1980); 
K. G. Begeman, A. H. Broeils, and R. H. Sanders, {\it Mon. Not. R. Astron. Soc.} {\bf 249}, 523 (1991); 
R. P. Olling and M. R. Merrifield, {\it Mon. Not. R. Astron. Soc.} {\bf 311}, 361 (2000). 

\bibitem{evidc}
M. Fich and S. Tremaine, {\it Annu. Rev. Astron. Astrophys.} {\bf 29}, 409 (1991).

\bibitem{WMAP}
WMAP Collab., D. N. Spergel et al., {\tt astro-ph/0603449} (2006).

\bibitem{susydm}
G. Jungman, M. Kamionkowski, and K. Griest, {\it Phys. Rep.} {\bf 267}, 195 (1996);
G. Bertone, D. Hooper and J. Silk, {\it Phys. Rep.} {\bf 405}, 279 (2005), {\tt hep-ph/0404175}.

\bibitem{detaa}
M. W. Goodman and E. Witten, {\it Phys. Rev.} {\bf D 31}, 3059 (1985);
I. Wassermann, {\it Phys. Rev.} {\bf D 33}, 2071 (1986);
K. Griest, {\it Phys. Rev.} {\bf D 38}, 2357 (1988).

\bibitem{detab}
A. K. Drukier, K. Freese, and D. N. Spergel, {\it Phys. Rev.} {\bf D 33}, 3495 (1986).

\bibitem{detac}
K. Freese, J. Frieman, and A. Gould, {\it Phys. Rev.} {\bf D 37}, 3388 (1988).

\bibitem{modela}
J. F. Navarro, C. S. Frenk, and S. D. M. White, {\it Astrophys. J.} {\bf 462}, 563 (1996);
A. V. Kravtsov {\it et al.}, {\it Astrophys. J.} {\bf 502}, 48 (1998);
B. Moore {\it et al.}, {\it Mon. Not. R. Astron. Soc.} {\bf 310}, 1147 (1999).

\bibitem{modelb}
N. W. Evans, {\it Mon. Not. R. Astron. Soc.} {\bf 260}, 191 (1993), and {\bf 267}, 333 (1994);
N. W. Evans, C. M. Carollo, and P. T. de Zeeuw, {\it Mon. Not. R. Astron. Soc.} {\bf 318}, 1131 (2000). 

\bibitem{modelc}
P. Sikivie, I. I. Tkachev, and Y. Wang, {\it Phys. Rev.} {\bf D 56}, 1863 (1997).

\bibitem{modeld}
M. Kamionkowski and A. Kinkhabwala, {\it Phys. Rev.} {\bf D 57}, 3256 (1998).

\bibitem{annualab}
P. Belli {\it et al.}, {\it Phys. Rev.} {\bf D 61}, 023512 (2000);
A. M. Green, {\it Phys. Rev.} {\bf D 63}, 043005 (2001).

\bibitem{annualaa}
D. N. Spergel, {\it Phys. Rev.} {\bf D 37}, 1353 (1988).

\bibitem{DAMA}
R. Bernabei {\it et al.}, {\it Phys. Lett.} {\bf B 480}, 23 (2000),
{\tt astro-ph/0305542} (2003), and {\tt astro-ph/0311046} (2003).

\bibitem{CDMS}
See {\tt http://cdms.berkeley.edu/}.

\bibitem{CRESST}
See {\tt http://www.cresst.de/}.

\bibitem{EDELWEISS}
See {\tt http://edelweiss.in2p3.fr/index\_newe.html}.

\bibitem{CDMSbound} 
CDMS Collab., D. S. Akerib {\it et al.}, {\it Phys. Rev. Lett.} {\bf 96}, 011302(2006),
{\tt astro-ph/0509259}.

\bibitem{gelmini}
P. Gondolo and G. Gelmini, {\it Phys. Rev.} {\bf D 71}, 123520 (2005), {\tt hep-ph/0504010}.

\bibitem{inel}
D. Tucker-Smith and N. Weiner, {\it Phys. Rev.} {\bf D 72}, 063509 (2005), {\tt hep-ph/0402065}.

\bibitem{Green07}
A. M. Green, {\tt hep-ph/0703217} (2007).

\bibitem{FQa}
S. P. Ahlen {\it et al.}, {\it Phys. Lett.} {\bf B 195}, 603 (1987).

\bibitem{FQb}
J. Engel, {\it Phys. Lett.} {\bf B 264}, 114 (1991).

\bibitem{Ling04}
 {F. S. Ling, P. Sikivie, and S. Wick, {\it Phys. Rev.} {\bf D 70}, 123503 (2004),
  {\tt astro-ph/0405231}.}

\end{thebibliography}
\end{document}